\documentclass[pdflatex,sn-standardnature]{sn-jnl}
\usepackage{times}
\usepackage{gensymb}

\usepackage{ulem}
\jyear{2022}%

\usepackage{etoolbox}
\begin{document}
\title{Spherical symmetry in the kilonova AT2017gfo/GW170817}

\author*[1,2]{\fnm{Albert} \sur{Sneppen}}\email{a.sneppen@gmail.com}
\author[1,2]{\fnm{Darach} \sur{Watson}}

\author[3]{\fnm{Andreas} \sur{Bauswein}}

\author[3,4]{\fnm{Oliver} \sur{Just}}

\author[5]{\fnm{Rubina} \sur{Kotak}}

\author[6]{\fnm{Ehud} \sur{Nakar}}

\author[6]{\fnm{Dovi} \sur{Poznanski}}

\author[7]{\fnm{Stuart} \sur{Sim}}

\affil[1]{Cosmic Dawn Center (DAWN), Copenhagen, Denmark}
\affil[2]{Niels Bohr Institute, University of Copenhagen, Copenhagen, Denmark}
\affil[3]{GSI Helmholtzzentrum f\"ur Schwerionenforschung, Darmstadt, Germany}
\affil[4]{Astrophysical Big Bang Laboratory, RIKEN Cluster for Pioneering Research, Tokyo, Japan}
\affil[5]{Department of Physics \& Astronomy, University of Turku, Turku, Finland}
\affil[6]{School of Physics and Astronomy, Tel-Aviv University, Tel-Aviv, Israel}
\affil[7]{School of Mathematics and Physics, Astrophysics Research Centre, Queen's University Belfast, Belfast, United Kingdom}

\abstract{The mergers of neutron stars expel a heavy-element enriched fireball which can be observed as a kilonova\cite{Eichler1989,Barnes2013,Tanvir2013,Berger2013}. The kilonova's geometry is a key diagnostic of the merger and is dictated by the properties of ultra-dense matter and the energetics of the collapse to a black hole. Current hydrodynamical merger models typically show aspherical ejecta\cite{Hotokezaka2013,Bauswein2013,Rosswog2014}. Previously, Sr+ was identified in the spectrum \cite{Watson2019} of the the only well-studied kilonova\cite{Abbott2017b,Pian2017,Smartt2017} AT2017gfo\cite{Coulter2017}, associated with the gravitational wave event GW170817. Here we combine the strong Sr\(^+\) P~Cygni absorption-emission spectral feature and the blackbody nature of kilonova spectrum, to determine that the kilonova is highly spherical at early epochs. Line shape analysis combined with the known inclination angle of the source\cite{Mooley2022} also shows the same sphericity independently. We conclude that energy injection by radioactive decay is insufficient to make the ejecta spherical. A magnetar wind or jet from the black-hole disk could inject enough energy to induce a more spherical distribution in the overall ejecta, however an additional process seems necessary to make the element distribution uniform. 
}

\maketitle

The most prominent feature in the spectra of AT2017gfo is a very broad emission and absorption line, observed at roughly $1\,\mu$m and $0.8\mu$m respectively. This is a P~Cygni feature expected from a rapidly expanding plasma and is due principally to the Sr\(^+\) 4p\(^6\)4d---4p\(^6\)5p transitions\cite{Watson2019,Gillanders2022}. The recognition that the \(0.8-1.0\mu\)m feature was a P~Cygni profile allows the expansion velocity of the photosphere along the line of sight, \(v_\parallel\), to be determined accurately at each epoch. The constraint on \(v_\parallel\) is not very dependent on the specific line identification since the emission component of the P~Cygni is centred at rest. 

\begin{figure}
     \includegraphics[width=\linewidth]{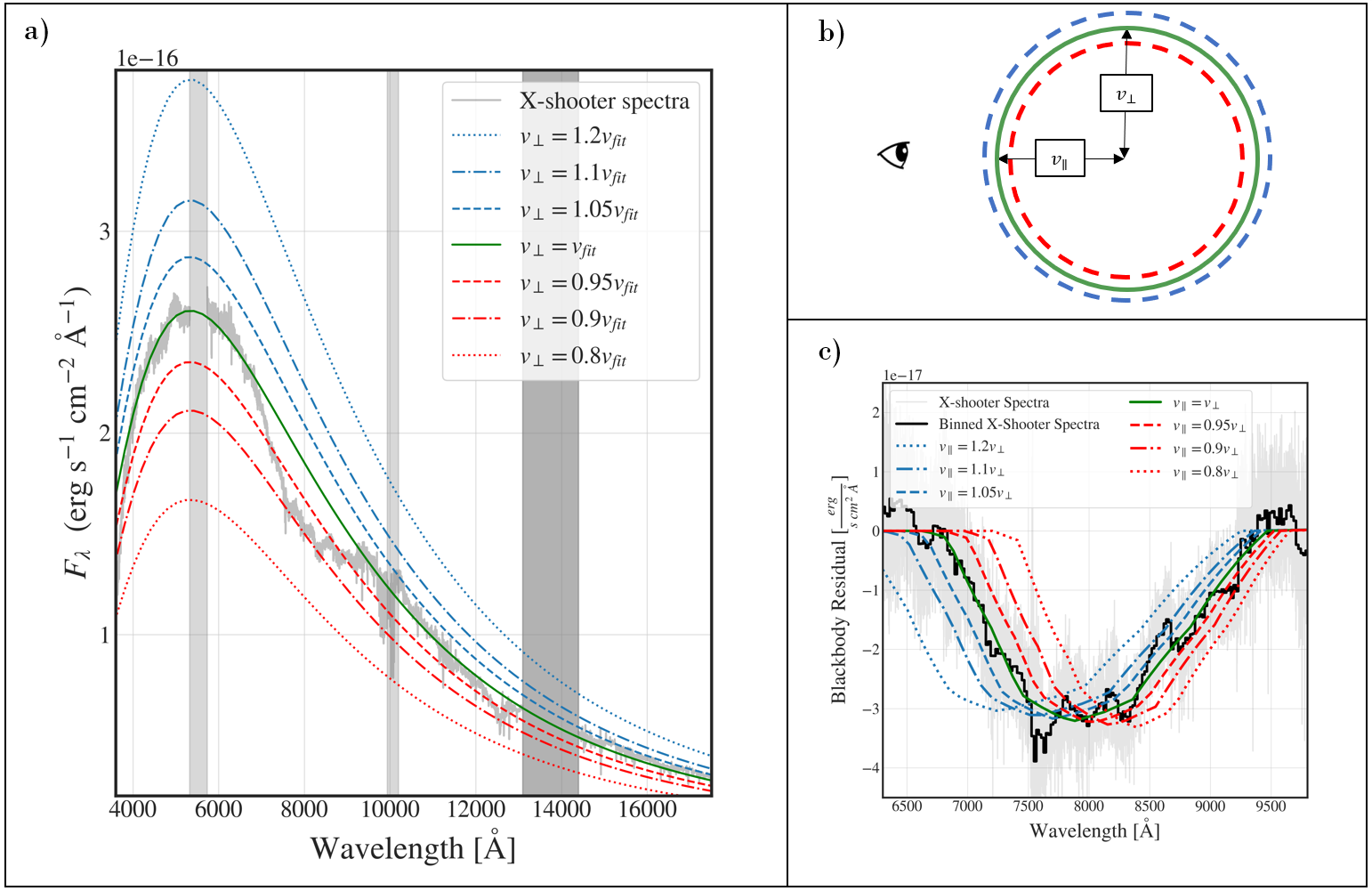}
     \caption{\textbf{Illustration of the expanding atmospheres method for the kilonova AT2017gfo.} a) Epoch~1 X-shooter spectrum with overlaid blackbody fits. The normalisation for each blackbody is set by the cross-sectional radius, which requires an {\it a priori} distance (here assuming the cosmological parameters from the \textit{Planck} mission\cite{Planck2018}) and cross-sectional velocity, $v_{\perp}$. Smaller (red), equal (green), or greater (blue) values of $v_{\perp}$ are shown, in comparison to the best-fit velocity, $v_{\parallel}$. b) Illustration of $v_{\parallel}$ and $v_{\perp}$ which are respectively set by the P~Cygni absorption feature and the blackbody normalisation. c) Zoom in of the Sr\,\textsc{ii} absorption component residual obtained by subtracting the X-shooter epoch~1 spectrum from the blackbody continuum fit. The P~Cygni profiles for $v_{\parallel}$ smaller (red), equal (green), or greater (blue) than $v_{\rm fit}$ are overlaid. The continuum measurement and absorption line fit yield velocities in tight agreement.}
    \label{fig:illus}
\end{figure}

The transverse velocity, \(v_\perp\), can also be inferred since we know the explosion time precisely from the gravitational wave signal\cite{Abbott2017a} and can measure the cross-sectional area at any epoch by comparing the observed, dereddened flux with the theoretical flux of the blackbody, predicted using the cosmological distance to the host galaxy. Thus, the combination of blackbody normalization and P~Cygni lines allows a unique measurement of the transverse to radial asymmetry of the kilonova photosphere (Fig.~\ref{fig:illus}). 

This technique when used to determine distances for core-collapse supernovae is referred to as the expanding photospheres method (EPM~\cite{Kirshner1974}, see Methods). In contrast to core-collapse supernovae, the correction to the photospheric radius due primarily to electron scattering, i.e.\ the dilution factor\cite{Eastman1996}, is expected to be negligible for kilonovae because of the much lower free electron number per unit mass and the much higher (bound-bound) opacity of the heavy elements that dominate their atmospheres\cite{Kasen2013} (see Methods).

In this analysis, we use the spectra taken between 1.4 and 5.4 days after the event\citep{Pian2017,Smartt2017} with the X-shooter spectrograph mounted on the Very Large Telescope at the European Southern Observatory. This is the best available spectral series of AT2017gfo. We follow the data reduction procedures outlined in ref.~\cite{Watson2019}. The spectra show clear temporal evolution starting from a blackbody shape at the earliest epochs with more prominent discrete spectral features appearing over time (see \ref{fig:X-shooter-spec-4}). We focus our analysis on spectral epochs 1 and 2 (1.4 and 2.4 days post-merger), as these spectra are in good agreement with coincident photometry, are well-described by a blackbody, and are not sensitive to the broad emission features at 1.5 and \(2.0\,\mu\)m (see Methods). 

We fit the spectrum at each epoch with a Planck function and a P~Cygni line profile associated with the Sr\(^+\) triplet at 1.0037, 1.0327, and 1.0915\,\(\mu\)m, to measure both the blackbody flux and velocity of expansion. The expansion velocity from the fit is \(v_\parallel\), while \(v_\perp\) was determined from the area of the photosphere inferred from the fitted blackbody flux, correcting for special relativistic and time-delay effects (see Methods) and using a  cosmological distance of $44.2 \pm 2.3$\,Mpc derived from the cosmic microwave background (CMB)\cite{Planck2018} with a Hubble constant \(H_0 = 67.36 \pm 0.54\) and a host galaxy peculiar velocity of $373 \pm 140$\,km/s\cite{Mukherjee2021}.

\begin{figure}
     \includegraphics[width=\linewidth]{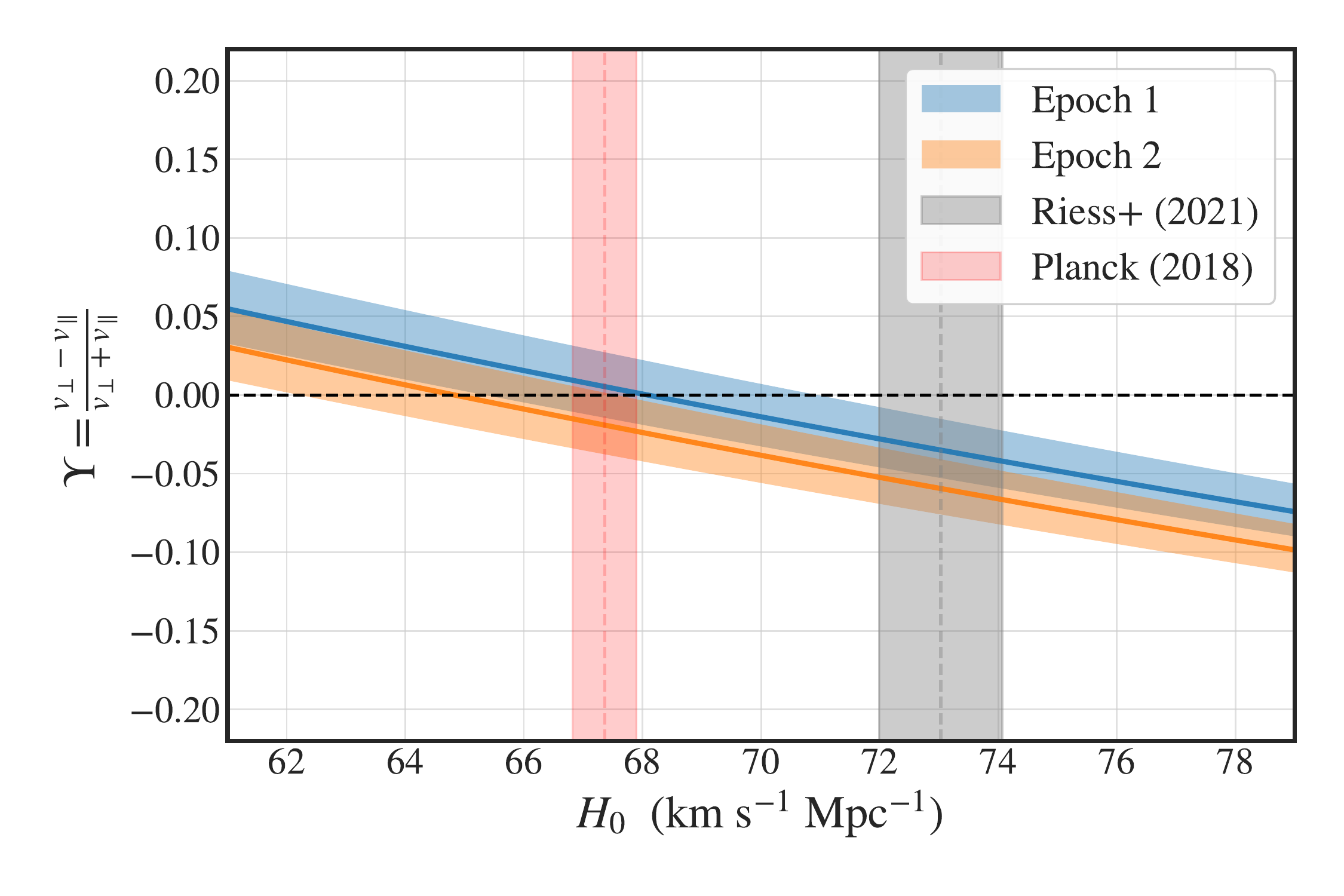}
     \caption{\textbf{Kilonova asymmetry index as a function of $H_0$}. The measurement of the asymmetry, $\Upsilon$, from the EPM method, and the Hubble constant are strongly degenerate, with higher $H_0$ implying more prolate ejecta. Shading indicates $1\sigma$ uncertainties with both early-universe (i.e.\ CMB\cite{Planck2018}) and late-universe estimates\cite{Riess2021} of $H_0$ suggesting that the kilonova is close to spherically symmetric, with less than 10\% variation between the velocities.}
    \label{fig:assym}
\end{figure}

Defining the zero-centred asymmetry index, \(\Upsilon = \frac{v_\perp-v_\parallel}{v_\perp+v_\parallel}\), we find \(\Upsilon = 0.00\pm0.02\) and \(\Upsilon = -0.02\pm0.02\) for the kilonova for epochs~1 and 2 respectively. The dominant uncertainty in $\Upsilon$ is the cosmologically-inferred distance to the source, whose uncertainty depends mostly on the peculiar velocity of the host galaxy, NGC\,4993 (see Methods). Using the local distance ladder \(H_0=73.03\pm1.04\)\cite{Riess2021} instead of the CMB value, does not change the high degree of spherical symmetry inferred (Fig.~\ref{fig:assym}), with an asymmetry \(\Upsilon = -0.04\pm0.03\) for epoch~1. We use the CMB-inferred \(H_0\) as our fiducial cosmology for the rest of this paper.

\begin{figure}
    \centering
    \includegraphics[width=0.9\linewidth]{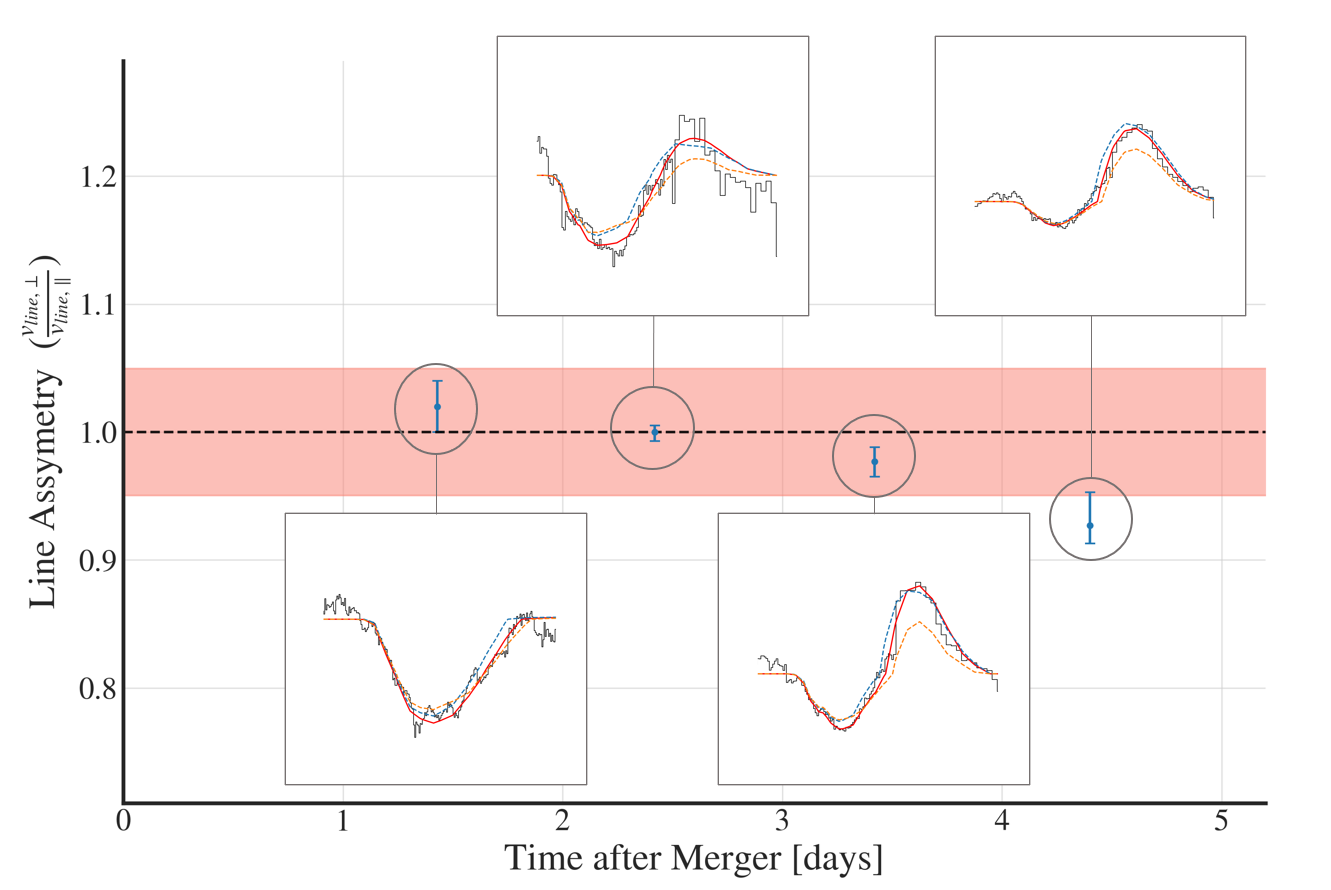}
    \caption{\textbf{Constraints on the spherical symmetry of the kilonova from the line shape.}  In every epoch the expansion is spherical to within a few percent. The error-bars shown are 68\% confidence limits. The red shading indicates the 1$\sigma$ sphericity constraints from the EPM method for epoch~1. Cutouts above and below the line show the residual to the blackbody fit around the $1\,\mu$m Sr\,\textsc{ii} feature for each epoch. The P~Cygni profile given fixed $v_{\parallel}$ is shown with varying ellipsoidal atmosphere shapes as follows: 5\% prolate (blue dashed line), spherical (red line), and 5\% oblate (orange dashed line). }
    \label{fig:line_shape_assym}
\end{figure}

We can also infer the sphericity of the ejecta by analysing the shape of the absorption line; for prolate or oblate geometries the absorption line will be skewed to higher or lower maximum velocities respectively. Typically, symmetry constraints from the shape of spectral lines are strongly degenerate with the viewing angle\cite{Hoeflich1996}. However, the comparison of the radio and optical astrometry of the counterparts to GW170817 provides good constraints on the inclination angle of the merger\cite{Mooley2022}. We therefore developed a P~Cygni profile model from expanding atmospheres with variable eccentricities (see Methods) with which we fit the P~Cygni profile. These independent line-shape sphericity constraints are shown in Fig.~\ref{fig:line_shape_assym}. For every epoch we find a line shape that is consistent with a completely spherical expansion to within a few percent. These line-shape constraints are independent of the EPM measurements and verify the spherical nature of the kilonova at early epochs.

The line shape offers further information, however, because it is sensitive to the angular distribution of the line-forming species. The sphericity implied by the line shape suggests a near spherically symmetric distribution of Sr\(^+\). This contrasts with the constraint on sphericity from the blackbody luminosity/line velocity method, which constrains the total opacity, not the opacity in any given element.

The measurement of such high degrees of sphericity are challenging to current magneto-hydrodynamic models of neutron star mergers. While a reasonable degree of sphericity in density and composition is in principle compatible with current models, it is not a generic outcome. Most models of the early dynamical ejecta feature an 
oblate density distribution due to the strong rotation of the system\cite{Hotokezaka2013,Bauswein2013} and a moderate pole-to-equator variation of the electron fraction $Y_e$ ($\Delta Y_e\sim 0.1$) because of more pronounced neutrino emission towards the poles\cite{Sekiguchi2016,Radice2018}, leading to a photosphere that is expected not to be spherical\cite{Just2015}.

Reaching spherical symmetry in the ejecta geometry through coincidences between the different mass ejection channels would not occur robustly and naturally over a large range of velocities. However, the spherical symmetry is observed to persist across several epochs with widely varying velocities, suggesting that such a coincidence between e.g.\ the dynamical and secular ejecta is unlikely.

The absorption line shape measurement also indicates that it is not just the total opacity, but also the Sr opacity, that is spherically symmetric. This makes a coincidence between a prolate density distribution exactly matched by an oblate specific opacity, improbable.

These results imply a uniformity in the matter opacity between the poles and equator, i.e.\ a small pole-to-equator variation of \(Y_e\). Current predictions of large differences between the ejected equatorial and polar matter compositions may be the result of uncertainties or incompleteness in the current models\cite{Foucart2018}. It is possible that the ejecta may be made more uniform in composition by neutrino flavor conversion processes\cite{Wu2017}, the physics of which are still not well-understood.

Regardless of the \(Y_e\) uniformity, the ejecta density itself needs to be made highly spherical. Adding a large injection of energy immediately after the merger could make the density distribution more spherical by puffing it up\cite{Rosswog2014}. We demonstrate using hydrodynamical models, however, that at least a few tens of MeV/nucleon are required to force a nearly spherical density distribution (see \ref{fig:injection}), ruling out radioactive heating as the source of this energy injection. We also find that very powerful heating alone (e.g.\ by magnetic reconnection) does not seem to significantly mix the equatorial and polar ejecta and generate a spherical composition and photosphere.

Energy may also be injected in an anisotropic fashion as a relativistic wind from the remnant neutron star or black hole by tapping the rotational energy of the system. Within the first few seconds of the explosion a polar outflow could be launched and  produce a rapidly-expanding balloon of high-\(Y_e\) material with low opacity, dominated by elements like Sr. This polar outflow, outpacing the equatorial ejecta, would expand sideways, covering the low-\(Y_e\) material and providing a near-spherical kilonova. Our hydrodynamical simulations (see \ref{fig:injection}) reveal that such a mechanism could result in a near-spherical distribution but seems to require tuning of the initial setup and the details of energy injection. 

No model is entirely satisfactory in explaining the remarkable spherical symmetry. Fundamentally, this finding is a surprise---including to the authors---as the sphericity of the ejecta differs from most current expectations. Type\,Ia supernovae offer an interesting comparison. Observations there also indicate a high degree of spherical symmetry in most cases\cite{Soker2019}, which is generally challenging to reproduce in models\cite{Bulla2016,Collins2022}.

\begin{figure}
    \centering
    \includegraphics[width=\linewidth]{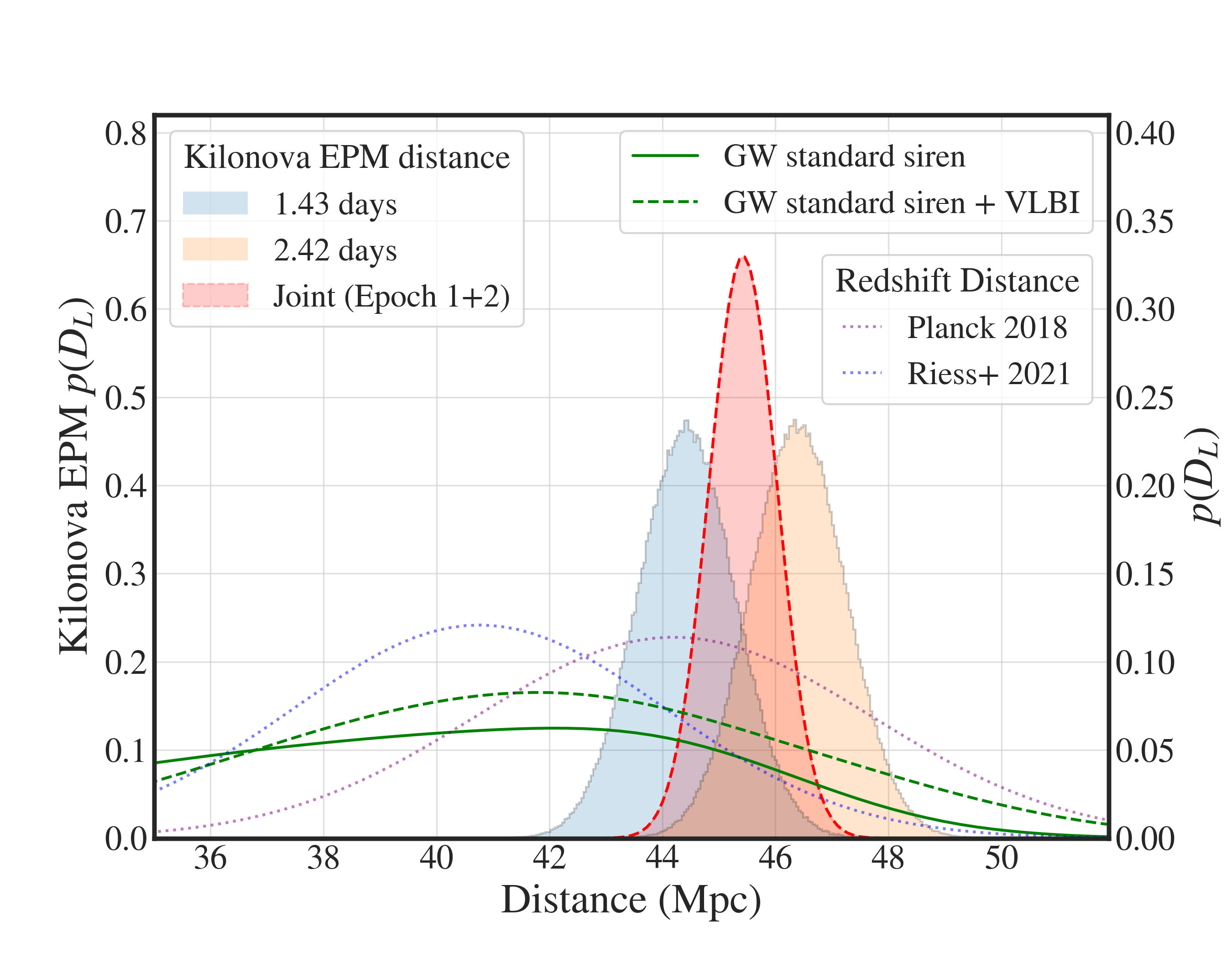} 
    \caption{\textbf{Posterior probability distributions for the luminosity distance to the kilonova AT2017gfo.} Distance estimates based on the kilonova expanding photospheres method for the spectra obtained at 1.43 (blue) and 2.42 (orange) days including sphericity corrections propagated from the line shape fits. The combined constraint is shown as a red histogram. The gravitational wave standard siren distance estimate \cite{Abbott2017b} is shown as a dashed green line. The standard siren estimate combined with very long baseline interferometry radio inclination angle data \cite{Mooley2018,Hotokezaka2019} is shown as a solid green line. Cosmological redshift distances are also plotted for \(H_0=67.36\pm0.54\) \cite{Planck2018} (dotted purple) and \(H_0=73.03\pm1.04\) \cite{Riess2021} (dotted blue).}
    \label{fig:d_l}
\end{figure}

The complementary nature of the sphericity constraints allows us to measure the distance to the kilonova directly, independent of the cosmological distance ladder. Since the sphericity of the ejecta can be measured from the line shape alone, the emitting area of the photosphere can be inferred directly by modelling the velocity and shape of the absorption line, with the inclination angle known from the radio jet, precision astrometry, or the gravitational standard siren, or a combination of these. We have used this method with the best-constrained inclination angle of ref~\cite{Mooley2022} to determine the luminosity distance to the kilonova, and find \(D_L = 45.5\pm0.6\)\,Mpc for the combined constraints from epochs 1 and 2 (Fig.~\ref{fig:d_l}). This uncertainty is around 1.3\%, which includes the propagated uncertainties associated with dust extinction and flux calibration. Systematic uncertainties related to the modelling could exceed this value (see Methods). Our distance measurement to AT2017gfo/GW170817 is consistent with the standard siren distance from the gravitational waves\cite{Abbott2017b} and the distance inferred from the cosmological redshift\cite{Mukherjee2021} for either the CMB\cite{Planck2018} or local distance-ladder\cite{Riess2021} inferred \(H_0\) value.


\newpage

\clearpage



\backmatter


\section*{Methods}
\counterwithin{figure}{section}
\renewcommand{\thefigure}{Extended Data Figure\,\arabic{figure}}
\renewcommand{\figurename}{} 
\setcounter{figure}{0}

\subsection*{The expanding photospheres method}
The expanding photospheres method (EPM) is used to measure luminosity distances, $D_L$, typically to supernovae\citep{Baade1926,Kirshner1974}. But EPM can also be applied to kilonovae. Early follow-up spectroscopy of kilonovae yield constraints on continuum and lines, while the gravitational wave and gamma-ray signals provide independent constraints on the time and orientation of the merger, and the environments around kilonovae are less obscured than their dustier supernova counterparts \citep{Gall2017}. Ultimately, these factors allow EPM to provide precise and internally consistent estimates on the distance to AT2017gfo. 

EPM assumes a simplified model of the ejecta as a photosphere in homologous and spherically symmetric expansion. Under these assumptions the photosphere size follows directly from the expansion velocity and the time since explosion. For supernovae, estimating the time of explosion requires fitting over multiple epochs, but for kilonovae the precise timing is already known from the gravitational wave signal. This removes a free parameter and ensures that each epoch can yield an estimate of distance that is independent of every other epoch. In combination with the luminosity of the blackbody relative to the observed, dereddened flux, this yields an estimate of the luminosity distance. 

The wavelength-specific luminosity of a spherical blackbody is \(L_{\lambda}^{\rm BB} = 4 \pi R_{\rm ph}^2 \pi B(\lambda,T)\), where 
\(B(\lambda,T)\) is the Planck function with temperature, \(T\), at wavelength \(\lambda\), and \(R_{\rm ph}\) is the photospheric radius and follows from the expansion velocity and the time since explosion, i.e.\ \(R_{\rm ph} = v_{\rm ph} (t-t_e)\). When the blackbody is expanding relativistically with velocity \(\beta = v_{\rm ph}/c\), the form of the observed spectrum can be approximated with an effective observed temperature, \(\pi B(\lambda,T_{\rm eff})f(\beta)\). That is, the observed luminosity is modified by time-delay effects and the relativistic Doppler corrections for an optically thick expanding source:
\begin{align}
     \frac{L_{\lambda}^{\rm BB}}{4 \pi R_{\rm ph}^2} &= 2 \pi \int_\beta^1 \delta(\mu)^5 B(\lambda', T') \left(\frac{1-\beta}{1-\beta \mu}\right)^2 \mu d\mu \\
     &= 2 \pi \int_\beta^1 B(\lambda, T_{\rm eff}(\mu)) \left(\frac{1-\beta}{1-\beta \mu}\right)^2 \mu d\mu\ = f(\beta,\lambda)\;\pi B(\lambda,T_{\rm eff})
\end{align}
Here the primed form indicates the quantity in the emitted and co-moving frame, $\mu = \cos(\theta)$ and the relativistic Doppler correction, \(\delta(\mu) = \left(\Gamma(1-\beta\mu)\right)^{-1}\), (see ref~\cite{Ghisellini2013}) sets the shift from the emitted blackbody temperature, \(T_{\rm emi}\), to the observed effective temperature, \(T_{\rm eff} = \delta(\mu)T_{\rm emi}\). The term \(((1-\beta)/(1-\beta \mu))^2\) represents the geometrical effect that in order to arrive at the same time, light arriving from the limb must have been emitted earlier, when the sphere had a smaller surface area\cite{Rees1967}. The correction term \(f(\beta,\lambda)\) generally depends on the spectral shape, but for this analysis the variations with wavelength are below 2\% over the entire spectral range. Thus, the observed spectrum can be modelled to good accuracy with a single-temperature blackbody, i.e.\ a constant \(f(\beta)\). The complexity of the modelling can be further increased by including the temporal evolution in temperature and/or expansion velocity e.g.\ as determined in ref.~\cite{Drout2017}, but for the mild time-delays of the kilonova, these higher-order effects are small compared to our uncertainties.
The specific total luminosity inferred from observations is $L_{\lambda}^{\rm obs} = 4 \pi D_L^2 F_\lambda$, where \(F_\lambda\) is the wavelength specific flux. Equating the blackbody-inferred and observed luminosities yields a relation for the angular size, $\theta$, of the ejecta: 


\begin{align}
    \theta = \frac{2 R_{\rm ph}}{D_\theta} = (1+z)^2 
    \frac{2 R_{\rm ph}}{D_L} = 2 (1+z)^2 \sqrt{\frac{F_{\lambda}}{\pi B(\lambda,T_{\rm eff})f(\beta)}}
\end{align}

Here, $D_\theta$ is the angular diameter distance which is converted to luminosity distance using $D_\theta (1+z)^2 = D_L$. Given the angular size and physical extent of the photosphere, the luminosity distance can be inferred \citep{Gall2016}: 

\begin{equation}
    D_L = (1+z)^2 \frac{2 R_{\rm ph}}{\theta} =  R_{\rm ph} \sqrt{\frac{\pi B(\lambda,T_{\rm eff})f(\beta)}{F_\lambda }}
\end{equation}

Crucially, this estimate requires no calibration with known distances and is entirely independent of the cosmic distance ladder. We note that historically the application of EPM to supernovae (SNe) has required the introduction of a large `dilution factor' to correct for the increase of the photospheric radius relative to thermalisation depth \cite{Eastman1996,Dessart2005,Dessart2015}. This increase is driven by the large contribution of electron-scattering opacity to the total opacity within the photospheres of SNe envelopes. However, in kilonova ejecta the electron scattering opacity is small compared to the opacity of bound-bound transitions. The ratio is lower for two reasons. First, the opacity is higher by several orders of magnitude due to the large number of lines the complex valence structure of $r$-process elements give rise to \cite{Kasen2013}. Second, the kilonova ejecta are composed of typically neutral, singly-, or doubly-ionized heavy elements, so the number of free electrons per unit mass is $O(10^{2})$ smaller than the typical value for ionized hydrogen. 
Thus, electron opacity makes a very small contribution, as $r$-process lines dominate the opacity in kilonova atmospheres. 

Conversely, comparing this distance estimate with any previous inferred distances yields an estimate of the cross-sectional area, which can be compared to the line-of-sight velocity. Translating the cross-sectional area to a single cross-sectional velocity requires an assumption of axial symmetry.  Given the close to polar inclination inferred from the radio measurements of the jet \cite{Mooley2018}, this corresponds to an assumption of cylindrical symmetry in the plane of the disk. The slight difference between exactly polar and the VLBI inclination constraints yield less than 1 percent variation in inferred cross-sectional area and, while we account for this effect in the cross-section, it is inconsequential to the constraints presented in this analysis.   

An equivalent method to measure the asymmetry is simply by comparing the inferred Hubble constant. The overall luminosity is set by the cross-sectional radius (i.e.\ the perpendicular velocity, $v_{\perp}$), while the main constraint on $v_{\rm ph}$ is derived from the absorption feature (i.e.\ the velocity of the photosphere along the line-of-sight, $v_{\parallel}$). 
Thus, by assuming $H_0$ we can compare the line-of-sight and orthogonal velocities. 
This method and the EPM are essentially interchangeable. Notable, constraints on the kilonova's asymmetry from spectropolarimetry at 1.43\,days are consistent with a spherical geometry, but in comparison with the EPM framework are not very constraining \cite{Covino2017,Bulla2019}.

We define the zero-centered asymmetry index $\Upsilon = \frac{v_{\perp}-v_\parallel}{v_{\perp}+v_\parallel}$. Here $\Upsilon<0$ would be a prolate expansion, corresponding to lower velocities in the plane of the binary than along the jet. This comparison does not probe the asymmetry of the mass distributions deeper substructures -- it only constrains the asymmetry of the ejecta photosphere. However, it is worth noting that hydrodynamical merger-models of kilonovae, in contrast to core-collapse SNe, are typically equally or more anisotropic in their outer layers compared to their deeper substructures \cite{Hotokezaka2013}.  

\subsection*{P~Cygni modelling}
For the P~Cygni modelling, we assume, as proposed in ref.~\citep{Watson2019} that the identification of the \(\sim1\,\mu\)m feature is due to the Sr\,\textsc{ii} 4d--5p transitions at 1,032.7\,nm, 1,091.5\,nm and 1,003.7\,nm \citep{Gillanders2022}. We show that these lines modelled with the P~Cygni profile prescription described below, describe the data well. We set the relative strengths of the lines as in ref.~\cite{Watson2019}. We tested the effect of modelling the feature with a single emission line, allowing the rest emission wavelength to be a free parameter. Despite having an additional free parameter, this single-line fit is worse than the three-line Sr\(^+\) fit.
However, even such a single line still supports spherical expansion with a distance of \(D_L = 45.3\pm0.5\)\,Mpc and a asphericity value \(\Upsilon = 0.01\pm 0.02\) assuming our fiducial distance \citep{Planck2018}.

The P~Cygni profile is characteristic of expanding envelopes where the same spectral line yields both an emission peak near the rest wavelength and a blueshifted absorption feature \citep{Jeffery1990}. The peak is formed by true emission or by scattering into the line of sight, while absorption is due to scattering of photospheric photons out of the line of sight. As the latter is in the front of the ejecta this component is blueshifted. The P~Cygni profile is characterised by several properties of the kilonova atmosphere. The optical depth determines the strength of absorption and emission, while the velocity of the ejecta sets the wavelength of the absorption minimum. For this analysis we use the implementation of the P~Cygni profile based on the Elementary Supernova model,\cite{Jeffery1990} where the profile is expressed in terms of the rest wavelength, $\lambda_0$, the line optical depth, $\tau$, scaling velocities for the velocity dependence of $\tau$, $v_{\rm e}$, the photospheric velocity, $v_{\rm ph}$, and the maximum ejecta velocity, $v_{\rm max}$. 
This implementation does not include the relative population of the states in the transition (i.e.\ the source function). We therefore include a parameter describing the enhancement of the P~Cygni emission. While the enhancement parameter improves the fit to the line-shape, it is the absorption component that provides the sphericity constraints for epochs~1 and 2 (see \ref{fig:all_epochs}), so the inclusion of this additional parameter does not substantially impact our conclusions. For comparison we have also implemented the P~Cygni profile prescription for relativistically expanding atmosphere in \cite{Hutsemekers1990}. For the mildly relativistic velocities analysed in this paper, the difference these frameworks is inconsequential with maximally a shift in $\Upsilon$ of $-0.01$, thus both frameworks staying coherently consistent with spherical. 

\subsubsection*{Elliptical P~Cygni Modelling}

In addition to the parameterization above, which is intrinsically spherical, we have developed a P~Cygni profile model from expanding atmospheres with an additional and variable eccentricity, $e$. This allows an independent probe of the sphericity by fitting the line shape. While in general constraints from the shape of spectral lines on the asymmetry of ejecta is degenerate with the viewing angle \cite{Hoeflich1996}, in this case, the radio jet emission in combination with Hubble Space Telescope precision astrometry provides good constraints on the inclination angle of the merger, with \(\theta_{inc} = 22\degree\pm3\degree\) \cite{Mooley2018,Mooley2022}. Therefore, we can constrain the eccentricity of the ellipse by fitting the spectral line shape. 

Physically, this parameterization is identical to the P~Cygni profile in the Elementary Supernova, but assumes the photosphere has an elliptical shape. Thus, the contribution of the initial specific intensity and the source function is derived at each resonance plane integrated over the varying shapes of the expanding atmosphere. Note, this utilizes an identical functional dependence of the line optical depth with velocity, $\tau(v)$, for all angles, but varies effectively with angle because varying geometries and inclination angles yield different source functions, initial specific intensity and contributing resonance planes. 

\subsection*{Robustness of the spectral modelling and epoch selection}

\begin{figure}
    \includegraphics[width=\linewidth]{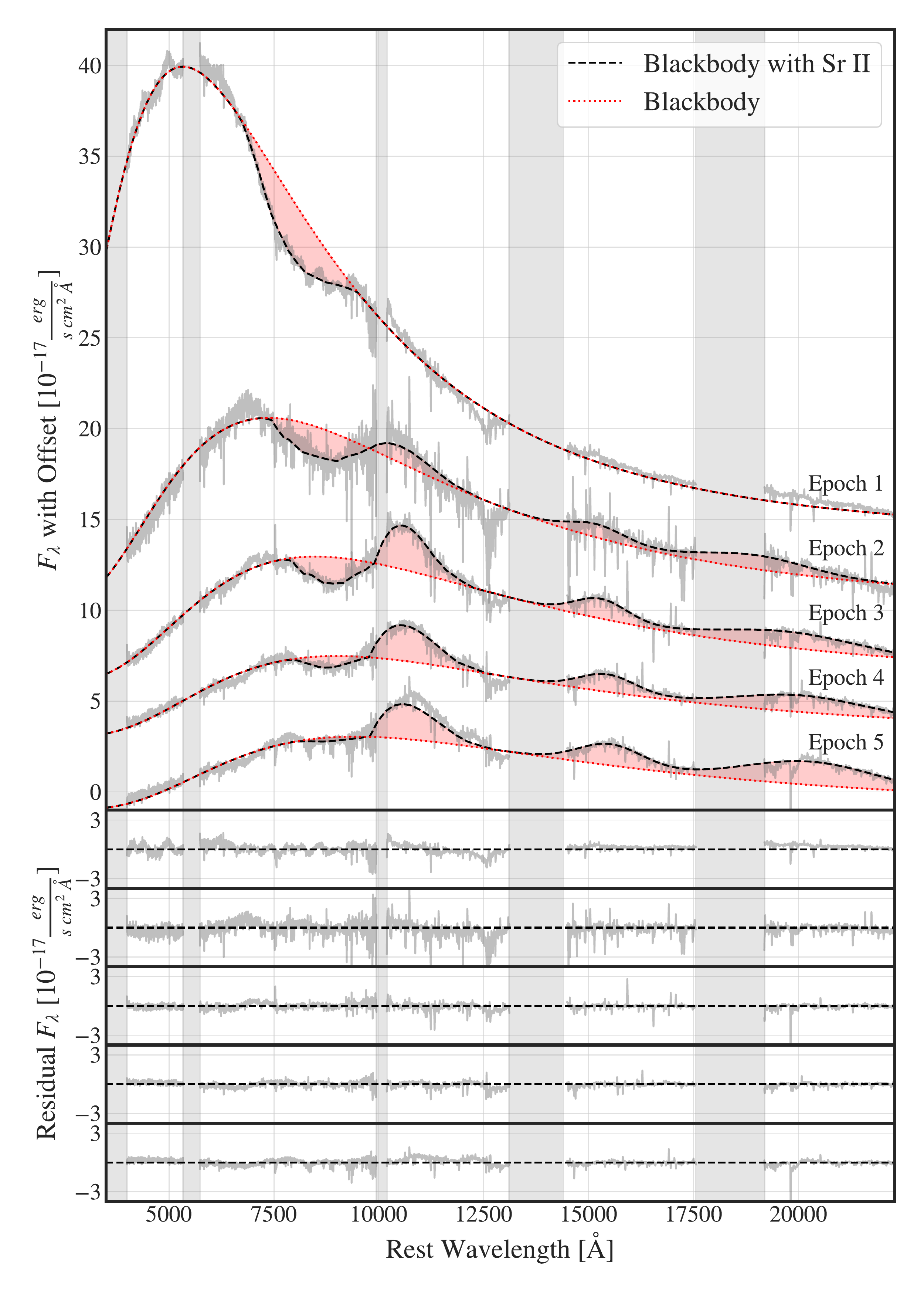}
    \caption{\textbf{Spectral series of AT2017gfo 1.4–5.4 days after the merger.} Spectra are from the VLT/X-shooter spectrograph (grey), with best fit shown with a dashed black line, and the blackbody-only component indicated with a red dotted line and deviations from the blackbody with pink fill. Grey-shaded regions were  not included in the fits. Darker shaded bars indicate telluric regions; light grey indicates overlapping noisy regions between the UVB, VIS and NIR arms of the spectrograph.}
    \label{fig:X-shooter-spec-4}
\end{figure}

Previous studies have established that the earliest epochs of AT2017gfo are well-approximated by a blackbody spectrum \citep{Malesani2017,Drout2017,Shappee2017}. With time, the spectrum increasingly deviates from a blackbody with progressively stronger and more numerous spectral components. 
We do not interpret the origin of these lines here. The resultant fit and spectrum for each of the first 5 epochs can be seen in \ref{fig:X-shooter-spec-4}.

Ultimately, for inferring distances or computing the asphericity index, the statistical uncertainty on the inferred photospheric velocity and the blackbody normalization is less than 1\% and thus subdominant to systematic uncertainties. These statistical uncertainties are tight due to the high velocities and the heavy-element dominated composition of kilonova ejecta, providing both a broad and well-constrained line and a remarkably Planck-like blackbody. For the distance determination the majority of the uncertainty resides in the flux calibration and dust extinction, while the peculiar velocity error dominates the uncertainty in estimating the asphericity.



For later epochs, additional complex components emerge, including the increasing prominence of the emission component of the \(\sim1\,\mu\)m Sr\(^+\) features, an emission feature at $\sim$1.4$\mu$m, clearly observed in the \textit{Hubble Space Telescope} spectra from the 5th and 10th days post-merger\cite{Tanvir2017}, a P~Cygni feature from at least the fourth day onwards at $\sim$0.75$\mu$m, and an emission component at $\sim$1.2$\mu$m from at least the 5th day onwards. While the models for the earliest epochs are largely independent of these additional spectral components, the modelling for later epochs is affected by whether and how these features are accounted for. We show in \ref{fig:all_epochs} the effect of including components in the modelling, or excluding spectral regions containing these components, to illustrate the sensitivity of the models to these effects. We still get agreement on the distances to order 10\% in all of the first five epochs, regardless of whether we include these components or not. More sophisticated modelling would almost certainly yield much better constraints. However, we use this analysis to show the robustness of our conclusions for the first two epochs. 
We emphasise that all epochs still support near-spherical expansion, as these are in agreement with the distance inferred from the cosmological redshift\cite{Mukherjee2021} for either the CMB\cite{Planck2018} or local distance-ladder\cite{Riess2021} inferred \(H_0\) value, while the typically larger distances in the latest epochs may hint of an increasingly prolate expansion as corroborated by the line-shape constraints in Fig. \ref{fig:line_shape_assym}. The spread in asymmetry values derived from the full model fit to epochs 1--3 is $\sigma_\Upsilon = 0.02$. As the constraint from each epoch is entirely independent, this provides corroborating evidence that the statistical uncertainty derived from sampling the posterior distribution is representative. Therefore, we also provide constraints from all epochs 1--5 to illustrate the generic robustness of the EPM framework applied to kilonovae.

\subsection*{Sr\(^+\) line as an indicator of photospheric velocity}

A critical assumption for any determined distance or asphericity is that the Sr\(^+\) lines are an accurate indicator of the photospheric velocity. This assumption is validated observationally in two separate respects. First, while a high optical depth of a line may detach it from the photosphere, the optical depth of Sr\(^+\) lines over all epochs analysed is moderate, yielding maximally a 20\% reduction in flux, with $\tau < 1$ indicated from the line fits. At these moderate optical depths the P~Cygni model is expected to trace the photospheric position \cite{Sim2017}. Second, the observed line-forming region continuously recedes deeper into the ejecta over time, which is why the distance remains consistent across multiple epochs despite the decreasing observed velocity. If Sr\(^+\) was detached from the photosphere, than as the photosphere recedes deeper into the ejecta, the velocity of the line-forming region and the photosphere would not naturally trace each other over time.

\subsection*{Systematic error due to line-blending}
Another potential concern is the effect of line-blending biasing the inferred photospheric velocity. The spectra do not require additional lines and while one can increase the complexity with a multitude of lines, this is unlikely to be a major concern for several reasons. First, it would require that the blending lines matched the Sr\(^+\) lines in strength over time, despite the major changes in density and temperature at each epoch. Second, since we get a similar distance and symmetry for each epoch, this hypothesis requires these lines to mimic a Sr\(^+\) P~Cygni profile, coherently shifting in wavelength and receding deeper within the ejecta at exactly the rate predicted by the blackbody continuum. However, to explore robustness and quantify the magnitude of the possible systematic error in such a line-blended case, we mapped the change in the inferred asymmetry that an additional P~Cygni line of freely varying $\tau$ would induce as a function of the central wavelength of the blending line. This reaches a maximum bias at about 9,900\,\AA, where it shifts the asymmetry index by $0.02$ and $0.03$ respectively for epochs 1 and 2. This is within our uncertainty margin, but we do not include it in our estimate of the error budget, because such a strong blending line is unlikely based on the arguments made above. Adding additional blending lines, while naturally increasing the uncertainty due to additional free parameters, does not increase the asymmetry shift more than adding a single additional line feature.


\subsection*{Distance estimates and potential constraints on \(H_0\)}

\begin{figure}
    \includegraphics[width=0.9\linewidth]{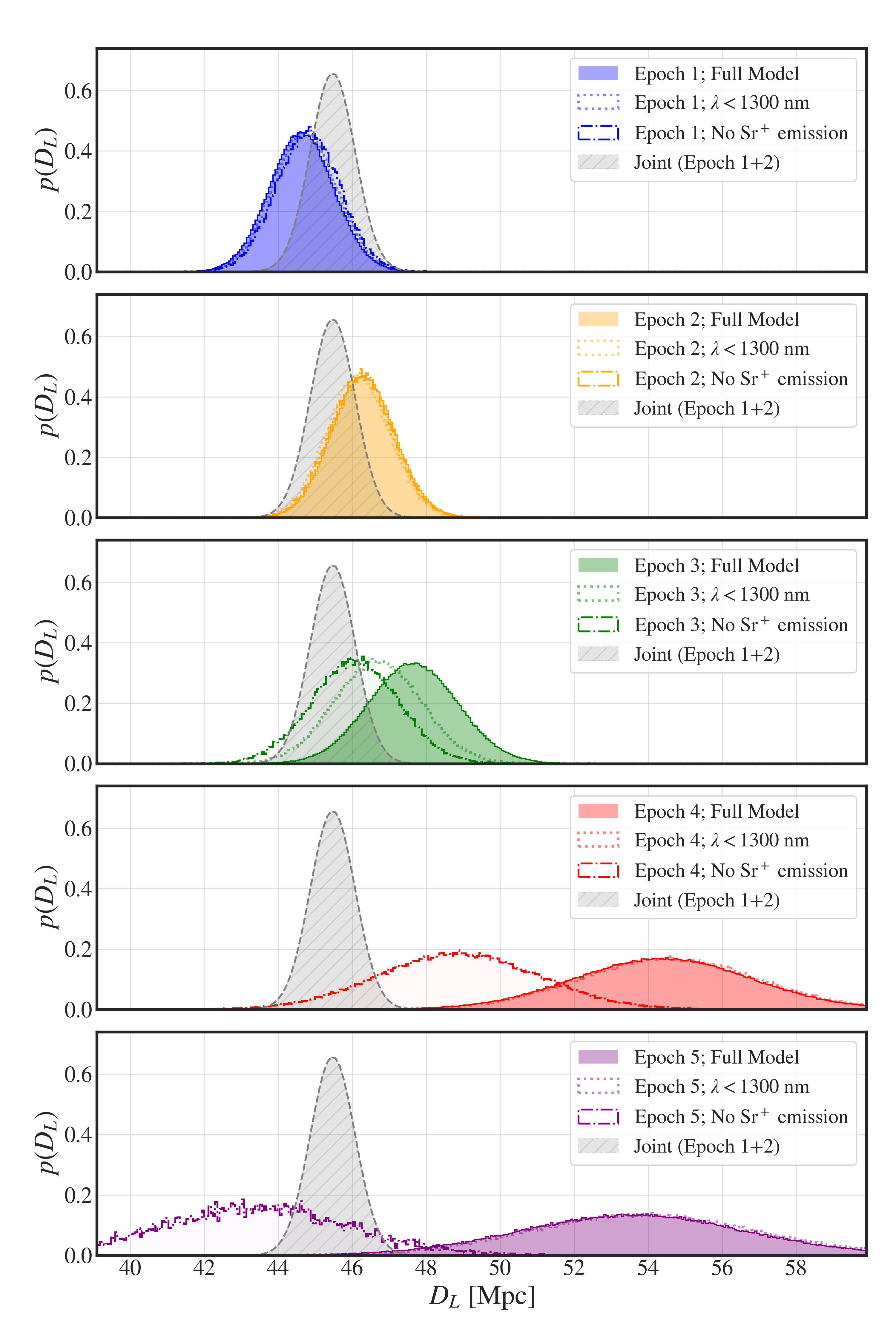} 
    \caption{\textbf{Posterior probability distributions of the luminosity distance to the kilonova AT2017gfo from epochs 1--5.} Our distance estimates based on the kilonova EPM for the spectra obtained at 1.43, 2.42, 3.41, 4.40 and 5.40 days are shown in blue, yellow, green, red, and purple histograms respectively. Filled histograms represent the full model including the blackbody continuum, Sr P~Cygni, and two NIR Gaussian emission lines. Dotted histograms indicate constraints from excluding all data with wavelengths longer than 1300\,nm, showing that the inclusion of the NIR Gaussian emission features do not bias the full model significantly. The dash-dotted histograms are for fits excluding the parts of the spectra with the Sr\(^+\) emission line. Distances derived from every epoch are consistent with the distances inferred from the GW standard siren plus VLBI constraints\cite{Hotokezaka2019}, however the distances inferred from epochs 3--5 are sensitive to the modelling of the $1\mu$m emission feature. In contrast, the data from epochs 1 and 2 provide robust, tight statistical uncertainties, with no large systematic variation between different models for emission components.}
    \label{fig:all_epochs}
\end{figure}

Assuming spherical symmetry, the EPM framework yields independent estimates for the luminosity distance from each epoch. The strict assumption of perfect spherical symmetry can be relaxed by modelling and propagating the symmetry measurements based on the shape of the absorption line, shown in Fig.~\ref{fig:line_shape_assym}. That is to say determining the cross-sectional radius, $R_{\rm \perp} = v_{\rm \perp} (t-t_e) = v_{\rm ph} (t-t_e) \left[\frac{v_{\rm \perp}}{v_{\rm ph}}\right]$, where the ratio in the square bracket follows directly from the line-shape fit, which yields the posterior distance distributions shown in Fig.~\ref{fig:d_l} and \ref{fig:all_epochs}.

\begin{figure}
    \centering
    \includegraphics[width=\linewidth]{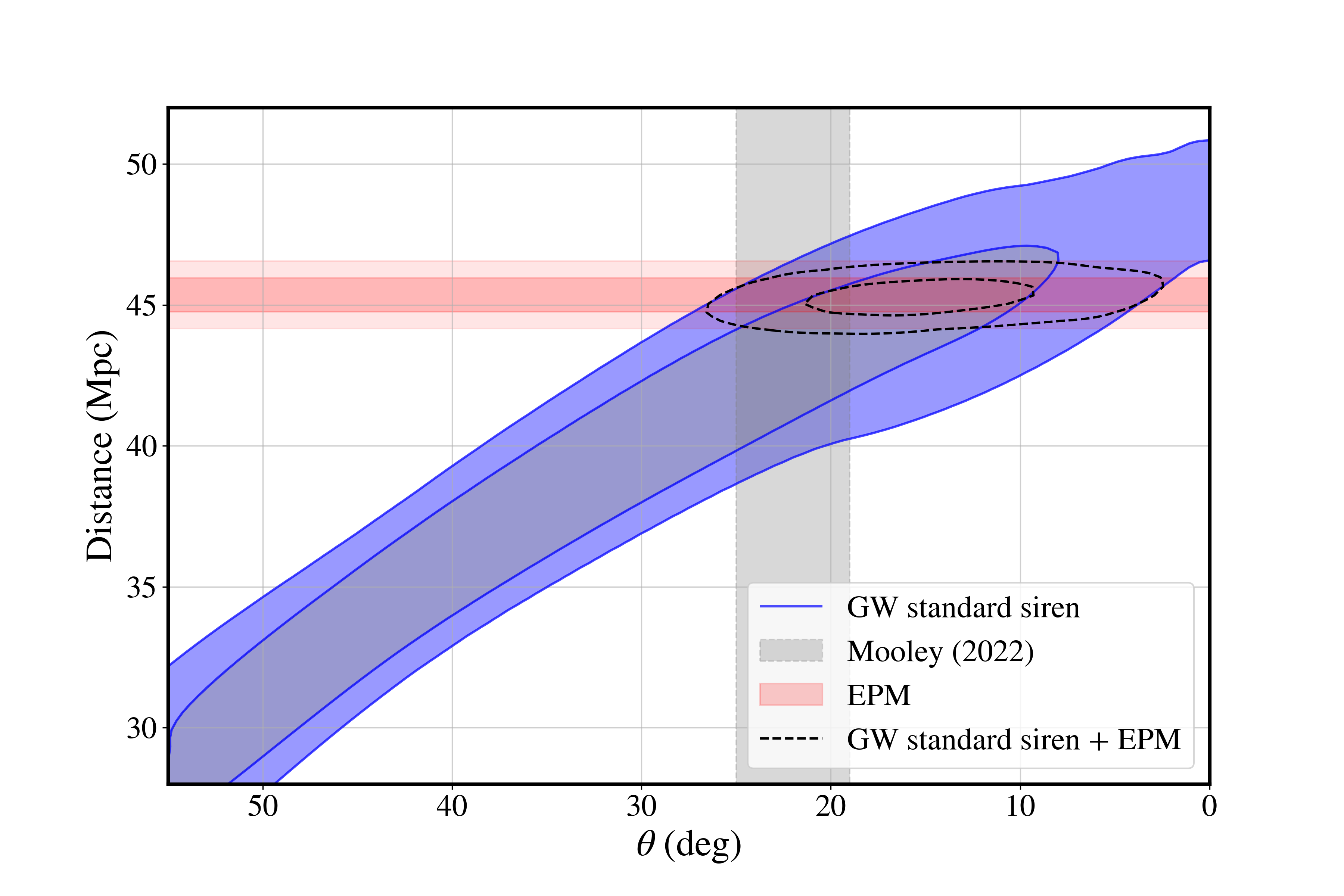}
    \caption{\textbf{Comparison of the inclination angle and luminosity distance to AT2017gfo compared to the inclination angle constraint from VLBI jet measurements.} The $1\sigma$ and $2\sigma$ constraints (dashed contours) from the combined EPM (red) and gravitational wave standard siren volumetric (blue) priors yield a tight constraint on the inclination angle, in close agreement with $1\sigma$ constraints from VLBI measurements and Hubble Space Telescope precision astrometry  (grey shading\cite{Mooley2022}).}
    \label{fig:inclination}
\end{figure}

The luminosity distances derived for the early epochs are consistent with previous estimates from the gravitational wave standard siren method \cite{Abbott2017b,Hotokezaka2019}, and from the cosmological redshift\cite{Mukherjee2021}, using either the \(H_0\) value inferred from the CMB\cite{Planck2018} or local distance-ladder\cite{Riess2021} methods. This distance is also useful because we can combine our distance measurement with the gravitational wave standard siren data\cite{Abbott2017b} to infer the angle of inclination of the merger plane (\ref{fig:inclination}) to be \(15\degree^{+6\degree}_{-5\degree}\) from the axis perpendicular to the plane. This is consistent with the inclination inferred from the radio measurements of the jet associated with the event and \textit{Hubble Space Telescope} precision astrometry, which yields \(22\degree\pm3\degree\)\cite{Mooley2022}.


Combining the EPM-estimated distance to the kilonova with the cosmological redshift of the host galaxy, can yield independent constraints on the Hubble constant, $H_0$, either assuming the ejecta is spherical or measuring the sphericity. To good approximation for $z \ll 1$, Hubble’s law gives the luminosity distance (with $q_0 = -0.53$ for standard cosmological parameters):

\begin{equation}
D_L \approx \frac{c z_{\rm cosmic}}{H_0} \left(1+\frac{1-q_0}{2}z_{\rm cosmic} \right)
\end{equation}
Here $z_{\rm cosmic}$ is the recession velocity due to the Hubble flow, i.e.\ correcting the observed redshift with respect to the CMB restframe, \(z_{\rm CMB}\) for the peculiar velocity \(z_{\rm pec}\). The recession velocity of the host-galaxy group is well-constrained $z_{\rm CMB} = 0.01110 \pm 0.00024$, but the peculiar velocity is harder to constrain \citep{Hjorth2017,Howlett2020,Nicolaou2020}. The most recent analysis, using a statistical reconstruction method which estimates large scale velocity flow using the Bayesian Origins Reconstruction from Galaxies (BORG), determines $z_{\rm pec} = 0.00124 \pm 0.00043$ \citep{Mukherjee2021}. This yields a cosmic recession velocity of \textbf{$z_{cosmic} = 0.00986 \pm 0.00049$}. The Hubble constant determined from our kilonova-EPM is $H_0= 65.6 \pm 3.4$\,km\,s$^{-1}$\,Mpc$^{-1}$ from the combined constraint of the first two epochs. The tight constraints on luminosity distances we find here means that the peculiar velocity is our dominant uncertainty when estimating $H_0$. However, with future detections, especially of somewhat more distant sources, this approach could yield precision constraints on \(H_0\) relatively quickly.


\subsection*{Numerical models of late-time energy injection}\label{sec:injectionmodels}

In order to test the ability of different types of late-time energy injection to explain the high degree of sphericity observed, we performed numerical simulations of toy-model ejecta configurations using the finite-volume hydrodynamics code AENUS-ALCAR \citep{Just2021}. We assume a similar physics input (initial ejecta distribution, special relativity, microphysical equation of state, axisymmetry) as in ref.\cite{Ito2021}, except that here we ignore neutrino interactions, which do not play a significant role during the late-time expansion. We do, however, adopt an initial composition (i.e. $Y_e$ profile) that approximates the outcome of simulations with detailed neutrino transport.

The scenario of local heating (representing, for example, energy release from radioactive decay or magnetic reconnection) is investigated by starting with a cold ejecta cloud of mass $0.04\,M_\odot$ and a slightly oblate geometry that is motivated by results for the dynamical ejecta obtained in neutron-star merger simulations \citep{Bauswein2013,Radice2018,Just2022}. The initial velocity of the cloud is given by $r/t_{\rm init}$ as a function of the distance from the center, $r$, where $t_{\rm init}=50\,$ms is the assumed post-merger time at which we start the simulation. The minimum (maximum) velocity of the bulk ejecta is $0.05\,c$ ($0.4\,c$). The density is distributed as $\propto r^{-n}$ with $n=3.5$ and a normalization factor depending on the polar angle $\theta$ as $0.25+\sin^3\theta$. The bulk ejecta are surrounded by a fast tail of very small mass, which decays with a power-law index much larger than $n$. As suggested by current state-of-the-art simulations of neutron-star mergers including neutrino interactions\citep{Fujibayashi2018, Ardevol-Pulpillo2019}, the electron fraction $Y_e$ in the ejecta is assumed to increase when going from equatorial to polar directions. Here we use $Y_e(\theta)=Y_{e,\rm min}+(Y_{e,\rm max}-Y_{e,\rm min})\cos\theta$ with $Y_{e,\rm min/max} = 0.2/0.4$. During the simulation, thermal energy is injected per unit of mass with a rate of

\begin{equation}
    Q_{\rm inj} = Q_0 \left(\frac{1}{2}-\frac{1}{\pi}\arctan\frac{t-1.3\,\mathrm{s}}{0.11\,\mathrm{s}}\right)^{1.3} \, ,
\end{equation}
inspired by radioactive heating in r-processed media\cite{Korobkin2012}. In this way, most of the energy is released within the first second of post-merger evolution. The factor $Q_0$ is chosen such that the total injected energy per baryon amounts to a fixed number of MeV. 
The models reveal that more than about 30\,MeV per baryon would be necessary to induce a nearly spherical density distribution at velocities greater than $0.2\,c$ (left panel of \ref{fig:injection}). Such high values exclude radioactive heating as a viable agent to sphericize the ejecta, because the maximum binding energy (obtained in $^{56}$Fe) is 8.8\,MeV and a large fraction of radioactive energy is lost to the emission of neutrinos, which do not thermalize. Energy dissipation through magnetic reconnection in highly magnetized media could possibly be powerful enough to achieve such high heating rates, but it is unclear whether magnetic fields are distributed homogeneously enough in the ejecta to end up with a spherical density distribution. An additional problem of the local-heating scenario remains its inability to remove (or reduce) large-scale composition gradients, most importantly the pole-to-equator variation of $Y_e$.

Another possibility is that the central (black hole or neutron-star) remnant injects energy into the expanding outflow through a relativistic wind due to, e.g.\ the Blandford-Znajek process\cite{Blandford1977} or magnetar spindown emission\cite{Thompson2004,Metzger2018}. In contrast to the local heating scenario, the injection of energy into a finite region of the ejecta opens up the possibility of lateral mixing. Material near the polar axis (which is typically less neutron rich and therefore more likely to contain substantial amounts of strontium) can outpace slower, equatorial material, allowing it to expand into equatorial directions that would otherwise be occupied by neutron-rich, strontium-poor material.
We test this scenario using a prolate ejecta cloud with a mass of $\sim 0.065\,M_\odot$. The initial configuration is the same as before, except that the power-law index, $n$, now smoothly increases with polar angle from 3.5 at the equator to 2 at the pole, the maximum velocity increases in a similar fashion from $0.1\,c$ to $0.22\,c$, and now $Y_{e,\rm min/max} = 0.15/0.45$. The bulk ejecta are again surrounded by a fast tail. About 100\,ms after the initialization time of $t_{\rm init}=0.4\,$s, a relativistic wind is injected at the inner radial boundary for a duration of 100\,ms, which has a Lorentz factor of $5$, specific enthalpy of 20, luminosity (per hemisphere) of $10^{52}\,$erg\,s$^{-1}$, and a half-opening angle of $60^\circ$. Compared to the local-heating scenario, our wind injection model (right panel of \ref{fig:injection}) shows a distribution of (originally polar) high-$Y_e$ material that is more extended laterally, supporting the ability of late-time wind-injection scenarios to reduce composition anisotropies in the ejecta. The mass distribution, however, does not become perfectly spherically symmetric, particularly in the high-entropy, low-density bubble that is created by the wind.

In addition to the specific configurations described above, we tried various other choices for the initial ejecta distribution and the (heating and wind) injection parameters, keeping the total mass and injected energy in the range compatible with AT2017gfo (0.01--0.08\,$M_\odot$ and $\lesssim 10^{51}\,$erg, respectively). These calculations clearly show that those different mechanisms can help to generate a more spherical outflow, although none of our models produced the high degree of sphericity suggested by the P~Cygni profile in both the density and composition. Our set of investigated models is not exhaustive and thus late-time energy injection may well operate even more efficiently than in our simplified models to sphericize the ejecta at late times. However, the fine-tuning needed to obtain a fully spherical configuration through these mechanisms may point to a high degree of sphericity of the initial ejecta distribution.


\section*{Data Availability}
Work in this paper was based on observations made with European Space Observatory (ESO) telescopes at the Paranal Observatory under programmes 099.D-0382 (principal investigator E. Pian), 099.D-0622 (principal investigator P. D’Avanzo), 099.D-0376 (principal investigator S. J. Smartt) and 099.D-0191 (principal investigator A. Grado). The data are available at http://archive.eso.org.

\section*{Code Availability}
We use the implementation of the P Cygni profile in the Elementary Supernova from https://github.com/unoebauer/public-astro-tools with generalizations to include variable ellipticity, inclination angle and enhancement of emission. Extensions of P Cygni code and data analysis required for generating figures can be found at: https://github.com/Sneppen/Kilonova-analysis  

\makeatletter
\apptocmd{\thebibliography}{\global\c@NAT@ctr 30\relax}{}{}
\makeatother


\begin{thebibliography}{10}
\expandafter\ifx\csname url\endcsname\relax
  \def\url#1{\texttt{#1}}\fi
\expandafter\ifx\csname urlprefix\endcsname\relax\def\urlprefix{URL }\fi
\providecommand{\bibinfo}[2]{#2}

\bibitem{Eichler1989}
\bibinfo{author}{{Eichler}, D. et al.},
\newblock \bibinfo{title}{{Nucleosynthesis, neutrino
bursts and {\ensuremath{\gamma}}-rays from coalescing neutron stars}}.
\newblock \emph{\bibinfo{journal}{Nature}}
  \textbf{\bibinfo{volume}{340}}, \bibinfo{pages}{126-128}
  (\bibinfo{year}{1989}).

\bibitem{Barnes2013}
\bibinfo{author}{{Barnes}, J. et al.}, 
\newblock \bibinfo{title}{{Effect of a High Opacity on the Light Curves of Radioactively Powered Transients from Compact Object Mergers}}.
\newblock \emph{\bibinfo{journal}{Astrophys. J.}}
  \textbf{\bibinfo{volume}{775}}, \bibinfo{pages}{18}
  (\bibinfo{year}{2013}). 
 
\bibitem{Tanvir2013}
\bibinfo{author}{{Tanvir}, N. R. et al.},
\newblock \bibinfo{title}{{A 'kilonova' associated with the short-duration {\ensuremath{\gamma}}-ray burst GRB 130603B}}.
\newblock \emph{\bibinfo{journal}{Nature}}
  \textbf{\bibinfo{volume}{500}}, \bibinfo{pages}{547-549}
  (\bibinfo{year}{2013}).

\bibitem{Berger2013}
\bibinfo{author}{{Berger}, E. et al.},
\newblock \bibinfo{title}{{An r-process Kilonova Associated with the Short-hard GRB 130603B}}.
\newblock \emph{\bibinfo{journal}{Astrophys. J. Lett.}}
  \textbf{\bibinfo{volume}{774}}, \bibinfo{pages}{L23}
  (\bibinfo{year}{2013}). 
  
  \bibitem{Hotokezaka2013}
\bibinfo{author}{{Hotokezaka}, K. et al.},
\newblock \bibinfo{title}{{Mass ejection from the merger of binary neutron stars}}.
\newblock \emph{\bibinfo{journal}{Phys. Rev. D.}}
  \textbf{\bibinfo{volume}{87}}, \bibinfo{pages}{024001}
  (\bibinfo{year}{2013}).

\bibitem{Bauswein2013}
\bibinfo{author}{{Bauswein}, A. et al.},
\newblock \bibinfo{title}{{Systematics of Dynamical Mass Ejection, Nucleosynthesis, and Radioactively Powered Electromagnetic Signals from Neutron-star Mergers}}.
\newblock \emph{\bibinfo{journal}{Astrophys. J.}}
  \textbf{\bibinfo{volume}{773}}, \bibinfo{pages}{79}
  (\bibinfo{year}{2013}).
  
\bibitem{Rosswog2014}
\bibinfo{author}{{Rosswog}, S. et al.},
\newblock \bibinfo{title}{{The long-term evolution of neutron star merger remnants - I. The impact of r-process nucleosynthesis}}.
\newblock \emph{\bibinfo{journal}{Mon. Not. R. Astron. Soc.}}
  \textbf{\bibinfo{volume}{439}}, \bibinfo{pages}{744-756}
  (\bibinfo{year}{2014}).

\bibitem{Watson2019}
\bibinfo{author}{{Watson}, D. et al.},
\newblock \bibinfo{title}{{Identification of strontium in the merger of two neutron stars}}.
\newblock \emph{\bibinfo{journal}{Nature}}
  \textbf{\bibinfo{volume}{574}}, \bibinfo{pages}{497–500}
  (\bibinfo{year}{2019}).

\bibitem{Abbott2017b}
\bibinfo{author}{{Abbott}, B.~P. et al.}
\newblock \bibinfo{title}{{Gravitational Waves and Gamma-Rays from a Binary Neutron Star Merger: GW170817 and GRB 170817A}}.
\newblock \emph{\bibinfo{journal}{Astrophys. J. Lett.}}
  \textbf{\bibinfo{volume}{848}}, \bibinfo{pages}{L13}
  (\bibinfo{year}{2017}).

\bibitem{Pian2017}
\bibinfo{author}{{Pian}, E. et al.}
\newblock \bibinfo{title}{{Spectroscopic identification of r-process nucleosynthesis in a double neutron-star merger}}.
\newblock \emph{\bibinfo{journal}{Nature}}
  \textbf{\bibinfo{volume}{551}}, \bibinfo{pages}{67-70}
  (\bibinfo{year}{2017}).

\bibitem{Smartt2017}
\bibinfo{author}{{Smartt}, S.~J. et al.}
\newblock \bibinfo{title}{{A kilonova as the electromagnetic counterpart to a gravitational-wave source.}}.
\newblock \emph{\bibinfo{journal}{Nature}}
  \textbf{\bibinfo{volume}{551}}, \bibinfo{pages}{75-79}
  (\bibinfo{year}{2017}).

\bibitem{Coulter2017}
\bibinfo{author}{{Coulter}, D. A. et al.},
\newblock \bibinfo{title}{{The optical counterpart to a gravitational wave source}}.
\newblock \emph{\bibinfo{journal}{Science}}
  \textbf{\bibinfo{volume}{358}}, \bibinfo{pages}{1556–1558}
  (\bibinfo{year}{2017}).
  
\bibitem{Mooley2022}
\bibinfo{author}{{Mooley}, K.~P. et al.},
\newblock \bibinfo{title}{{Optical superluminal motion measurement in the neutron-star merger GW170817}}.
\newblock \emph{\bibinfo{journal}{Nature}}
  \textbf{\bibinfo{volume}{610}}, \bibinfo{pages}{273-276}
  (\bibinfo{year}{2022}).

\bibitem{Gillanders2022}
\bibinfo{author}{{Gillanders}, J.~H. et al.},
\newblock \bibinfo{title}{{Modelling the spectra of the kilonova AT2017gfo - I. The photospheric epochs}}.
\newblock \emph{\bibinfo{journal}{Mon. Not. R. Astron. Soc.}}
  \textbf{\bibinfo{volume}{515}}, \bibinfo{pages}{631-651}
  (\bibinfo{year}{2022}).

\bibitem{Abbott2017a}
\bibinfo{author}{{Abbott}, B.~P. et al.}, 
\newblock \bibinfo{title}{{A gravitational-wave standard siren measurement of the Hubble constant}}.
\newblock \emph{\bibinfo{journal}{Nature}} \textbf{\bibinfo{volume}{551}},
  \bibinfo{pages}{85--88} (\bibinfo{year}{2017}).

\bibitem{Kirshner1974}
\bibinfo{author}{{Kirshner}, R.~P. et al.},
\newblock \bibinfo{title}{{Distances to extragalactic supernovae}}.
\newblock \emph{\bibinfo{journal}{Astrophys. J.}}
  \textbf{\bibinfo{volume}{228}}, \bibinfo{pages}{359}
  (\bibinfo{year}{1926}).

\bibitem{Eastman1996}
\bibinfo{author}{{Eastman}, R.~G. et al.}, 
\newblock \bibinfo{title}{{The Atmospheres of Type II Supernovae and the Expanding Photosphere Method}}.
\newblock \emph{\bibinfo{journal}{Astrophys. J.}} \textbf{\bibinfo{volume}{466}},
  \bibinfo{pages}{911} (\bibinfo{year}{1996}).  

\bibitem{Kasen2013}
\bibinfo{author}{{Kasen}, D. et al.}, 
\newblock \bibinfo{title}{{Opacities and Spectra of the r-process Ejecta from Neutron Star Mergers}}.
\newblock \emph{\bibinfo{journal}{Astrophys. J.}} \textbf{\bibinfo{volume}{774}},
  \bibinfo{pages}{25} (\bibinfo{year}{2013}).  

\bibitem{Mukherjee2021}
\bibinfo{author}{{Mukherjee}, S. et al.}, 
\newblock \bibinfo{title}{{Velocity correction for Hubble constant measurements from standard sirens}}.
\newblock \emph{\bibinfo{journal}{Astronomy 
and Astrophysics}} \textbf{\bibinfo{volume}{646}},
\bibinfo{pages}{A65} (\bibinfo{year}{2021}).  

\bibitem{Planck2018}
\bibinfo{author}{{Aghanim}, N. et al.}, 
\newblock \bibinfo{title}{{Planck 2018 results. VI. Cosmological parameters}}.
\newblock \emph{\bibinfo{journal}{arXiv e-prints}}, (\bibinfo{year}{2018}). 

\bibitem{Riess2021}
\bibinfo{author}{{Riess}, A.~G. et al.}, 
\newblock \bibinfo{title}{{A Comprehensive Measurement of the Local Value of the Hubble Constant with 1 km s$^{-1}$ Mpc$^{-1}$ Uncertainty from the Hubble Space Telescope and the SH0ES Team}}.
\newblock \emph{\bibinfo{journal}{Astrophys. J. Lett.}} \textbf{\bibinfo{volume}{934}},
\bibinfo{pages}{L7} (\bibinfo{year}{2022}).  

\bibitem{Hoeflich1996}
\bibinfo{author}{{Hoeflich}, I. et al.}, 
\newblock \bibinfo{title}{{Analysis of the Polarization and Flux Spectra of SN 1993J}}.
\newblock \emph{\bibinfo{journal}{Astrophys. J.}} \textbf{\bibinfo{volume}{459}},
\bibinfo{pages}{307} (\bibinfo{year}{1996}).

\bibitem{Sekiguchi2016}
\bibinfo{author}{{Sekiguchi}, Y. et al.}, 
\newblock \bibinfo{title}{{Dynamical mass ejection from the merger of asymmetric binary neutron stars: Radiation-hydrodynamics study in general relativity}}.
\newblock \emph{\bibinfo{journal}{Phys. Rev. D}} \textbf{\bibinfo{volume}{93}},
\bibinfo{pages}{124046} (\bibinfo{year}{2016}).

\bibitem{Radice2018}
\bibinfo{author}{{Radice}, D. et al.}, 
\newblock \bibinfo{title}{{Binary Neutron Star Mergers: Mass Ejection, Electromagnetic Counterparts, and Nucleosynthesis}}.
\newblock \emph{\bibinfo{journal}{Astrophys. J.}} \textbf{\bibinfo{volume}{869}},
\bibinfo{pages}{130} (\bibinfo{year}{2018}).

\bibitem{Just2015}
\bibinfo{author}{{Just}, O. et al.}, 
\newblock \bibinfo{title}{{Comprehensive nucleosynthesis analysis for ejecta of compact binary mergers}}.
\newblock \emph{\bibinfo{journal}{Mon. Not. R. Astron. Soc.}} \textbf{\bibinfo{volume}{448}},
\bibinfo{pages}{541-567} (\bibinfo{year}{2015}).

\bibitem{Foucart2018}
\bibinfo{author}{{Foucart}, F. et al.}, 
\newblock \bibinfo{title}{{Evaluating radiation transport errors in merger simulations using a Monte Carlo algorithm}}.
\newblock \emph{\bibinfo{journal}{Phys. Rev. D}} \textbf{\bibinfo{volume}{98}},
\bibinfo{pages}{063007} (\bibinfo{year}{2018}).

\bibitem{Wu2017}
\bibinfo{author}{{Wu}, M.~R.} \& \bibinfo{author}{{Tamborra}, I.}
\newblock \bibinfo{title}{{Fast neutrino conversions: Ubiquitous in compact binary merger remnants}}.
\newblock \emph{\bibinfo{journal}{Phys. Rev. D}} \textbf{\bibinfo{volume}{95}},
\bibinfo{pages}{103007} (\bibinfo{year}{2017}).

\bibitem{Soker2019}
\bibinfo{author}{{Soker}, N. et al.}
\newblock \bibinfo{title}{{Supernovae Ia in 2019 (review): A rising demand for spherical explosions}}.
\newblock \emph{\bibinfo{journal}{New Astronomy Reviews}} \textbf{\bibinfo{volume}{87}},
\bibinfo{pages}{101535} (\bibinfo{year}{2019}).

\bibitem{Bulla2016}
\bibinfo{author}{{Bulla}, M. et al.}
\newblock \bibinfo{title}{{Type Ia supernovae from violent mergers of carbon-oxygen white dwarfs: polarization signatures.}}.
\newblock \emph{\bibinfo{journal}{Mon. Not. R. Astron. Soc.}} \textbf{\bibinfo{volume}{455}},
\bibinfo{pages}{1060-1070} (\bibinfo{year}{2016}).


\bibitem{Collins2022}
\bibinfo{author}{{Collins}, S. et al.}
\newblock \bibinfo{title}{{Double detonations: variations in Type Ia supernovae due to different core and He shell masses - II: synthetic observables}}.
\newblock \emph{\bibinfo{journal}{Mon. Not. R. Astron. Soc.}} (\bibinfo{year}{2022}).

\end{thebibliography}

\begin{thebibliography}{10}\label{secondbib}


\bibitem{Baade1926}
\bibinfo{author}{{Baade}, W.},
\newblock \bibinfo{title}{{{\"U}ber eine M{\"o}glichkeit, die Pulsationstheorie der {\ensuremath{\delta}} Cephei-Ver{\"a}nderlichen zu pr{\"u}ten}}.
\newblock \emph{\bibinfo{journal}{Astronomische Nachrichten}}
  \textbf{\bibinfo{volume}{193}}, \bibinfo{pages}{27-36}
  (\bibinfo{year}{1974}).
  
\bibitem{Gall2017}
\bibinfo{author}{{Gall}, C. et al.},
\newblock \bibinfo{title}{{Lanthanides or Dust in Kilonovae: Lessons Learned from GW170817}}.
\newblock \emph{\bibinfo{journal}{Astrophys. J. Lett.}}
  \textbf{\bibinfo{volume}{849}}, \bibinfo{pages}{L29}
  (\bibinfo{year}{2017}).



\bibitem{Ghisellini2013}
\bibinfo{author}{{Ghisellini}, G.},
\newblock \bibinfo{title}{{Radiative Processes in High Energy Astrophysics}}.
\newblock 
  \textbf{\bibinfo{volume}{873}}
  (\bibinfo{year}{2013}).
 
\bibitem{Rees1967}
\bibinfo{author}{{Rees}, M.~J.},
\newblock \bibinfo{title}{{Studies in radio source structure-I. A relativistically expanding model for variable quasi-stellar radio sources}}.
\newblock \emph{\bibinfo{journal}{Mon. Not. R. Astron. Soc.}}
  \textbf{\bibinfo{volume}{135}}, \bibinfo{pages}{345}
  (\bibinfo{year}{1967}). 
    
\bibitem{Drout2017}
\bibinfo{author}{{Drout}, M.~R. et al.},
\newblock \bibinfo{title}{{Light curves of the neutron star merger GW170817/SSS17a: Implications for r-process nucleosynthesis}}.
\newblock \emph{\bibinfo{journal}{Science}}
  \textbf{\bibinfo{volume}{358}}, \bibinfo{pages}{1570-1574}
  (\bibinfo{year}{2017}).
  
\bibitem{Gall2016}
\bibinfo{author}{{Gall}, E. et al.},
\newblock \bibinfo{title}{{Applying the expanding photosphere and standardized candle methods to Type II-Plateau supernovae at cosmologically significant redshifts }}.
\newblock \emph{\bibinfo{journal}{Astronomy 
and Astrophysics}} \textbf{\bibinfo{volume}{592}}, \bibinfo{pages}{A129}
  (\bibinfo{year}{2016}).
  
\bibitem{Dessart2005}
\bibinfo{author}{{Dessart}, L.}  \& \bibinfo{author}{{Hillier}, D.~J.},
\newblock \bibinfo{title}{{Distance determinations using type II supernovae and the expanding photosphere method}}.
\newblock \emph{\bibinfo{journal}{Astronomy 
and Astrophysics}} \textbf{\bibinfo{volume}{439}}, \bibinfo{pages}{671-685}
  (\bibinfo{year}{2005}).
  
\bibitem{Dessart2015}
\bibinfo{author}{{Dessart}, L. et al.},
\newblock \bibinfo{title}{{Radiative-transfer models for supernovae IIb/Ib/Ic from binary-star progenitors}}.
\newblock \emph{\bibinfo{journal}{Mon. Not. R. Astron. Soc.}} \textbf{\bibinfo{volume}{453}}, \bibinfo{pages}{2189-2213}
  (\bibinfo{year}{2015}).
  
\bibitem{Mooley2018}
\bibinfo{author}{{Mooley}, K.~P. et al.},
\newblock \bibinfo{title}{{Superluminal motion of a relativistic jet in the neutron-star merger GW17081}}.
\newblock \emph{\bibinfo{journal}{Nature}} \textbf{\bibinfo{volume}{561}}, \bibinfo{pages}{355-359}
  (\bibinfo{year}{2018}).
  
\bibitem{Covino2017}
\bibinfo{author}{{Covino}, S. et al.}, 
\newblock \bibinfo{title}{{The unpolarized macronova associated with the gravitational wave event GW170817}}.
\newblock \emph{\bibinfo{journal}{Nature Astronomy}} \textbf{\bibinfo{volume}{1}},
\bibinfo{pages}{791-794} (\bibinfo{year}{2017}).

\bibitem{Bulla2019}
\bibinfo{author}{{Bulla}, M. et al.}, 
\newblock \bibinfo{title}{{The origin of polarization in kilonovae and the case of the gravitational-wave counterpart AT2017gfo}}.
\newblock \emph{\bibinfo{journal}{Nature Astronomy}} \textbf{\bibinfo{volume}{3}},
\bibinfo{pages}{99-106} (\bibinfo{year}{2019}).
  
\bibitem{Jeffery1990}
\bibinfo{author}{{Jeffery}, D.~J.}, \bibinfo{author}{{Branch}, D. (eds)},
\newblock \bibinfo{title}{{Analysis of Supernova Spectra}}. \textbf{\bibinfo{volume}{6}}, \bibinfo{pages}{149}
  (\bibinfo{year}{1990}).
  
\bibitem{Hutsemekers1990}
\bibinfo{author}{{Hutsemekers}, D.}  \& \bibinfo{author}{{Surdej}, J.},
\newblock \bibinfo{title}{{Formation of P Cygni Line Profiles in Relativistically Expanding Atmospheres.}}.
\newblock \emph{\bibinfo{journal}{Astrophys. J.}} \textbf{\bibinfo{volume}{361}}, \bibinfo{pages}{367}
  (\bibinfo{year}{1990}).
  
\bibitem{Malesani2017}
\bibinfo{author}{{Malesani}, D. et al.},
\newblock \bibinfo{title}{{LIGO/Virgo G298048: optical spectral energy distribution of SSS17a}}.
\newblock \emph{\bibinfo{journal}{GRB Coordinates Network}} \textbf{\bibinfo{volume}{21577}}, \bibinfo{pages}{1}
  (\bibinfo{year}{2017}).
  
\bibitem{Shappee2017}
\bibinfo{author}{{Shappee}, B.~J. et al.},
\newblock \bibinfo{title}{{Early spectra of the gravitational wave source GW170817: Evolution of a neutron star merger}}.
\newblock \emph{\bibinfo{journal}{Science}} \textbf{\bibinfo{volume}{358}}, \bibinfo{pages}{1574-1578}
  (\bibinfo{year}{2017}).
  
\bibitem{Tanvir2017}
\bibinfo{author}{{Tanvir}, N.~R. et al.},
\newblock \bibinfo{title}{{The Emergence of a Lanthanide-rich Kilonova Following the Merger of Two Neutron Stars}}.
\newblock \emph{\bibinfo{journal}{Astrophys. J. Lett.}} \textbf{\bibinfo{volume}{848}}, \bibinfo{pages}{L27}
  (\bibinfo{year}{2017}).
  
\bibitem{Sim2017}
\bibinfo{author}{{Sim}, S.~A.},
\newblock \bibinfo{title}{{Spectra of Supernovae During the Photospheric Phase}}.
\newblock \emph{\bibinfo{journal}{Handbook of Supernovae}} \bibinfo{pages}{769}
  (\bibinfo{year}{2017}).
  
\bibitem{Hotokezaka2019}
\bibinfo{author}{{Hotokezaka}, K. et al.},
\newblock \bibinfo{title}{{A Hubble constant measurement from superluminal motion of the jet in GW170817}}.
\newblock \emph{\bibinfo{journal}{Nature Astronomy}} \textbf{\bibinfo{volume}{3}}, \bibinfo{pages}{940-944}
  (\bibinfo{year}{2019}).

  
\bibitem{Hjorth2017}
\bibinfo{author}{{Hjorth}, J. et al.},
\newblock \bibinfo{title}{{The Distance to NGC 4993: The Host Galaxy of the Gravitational-wave Event GW170817}}.
\newblock \emph{\bibinfo{journal}{Astrophys. J. Lett.}} \textbf{\bibinfo{volume}{848}}, \bibinfo{pages}{L31}
  (\bibinfo{year}{2017}).
 
\bibitem{Howlett2020}
\bibinfo{author}{{Howlett}, C.}  \& \bibinfo{author}{{Davis}, T.~M.},
\newblock \bibinfo{title}{{Standard siren speeds: improving velocities in gravitational-wave measurements of H$_{0}$}}.
\newblock \emph{\bibinfo{journal}{Mon. Not. R. Astron. Soc.}} \textbf{\bibinfo{volume}{492}}, \bibinfo{pages}{3803-3815}
  (\bibinfo{year}{2020}).
  
\bibitem{Nicolaou2020}
\bibinfo{author}{{Nicolaou}, C. et al},
\newblock \bibinfo{title}{{Standard siren speeds: improving velocities in gravitational-wave measurements of H$_{0}$}}.
\newblock \emph{\bibinfo{journal}{Mon. Not. R. Astron. Soc.}} \textbf{\bibinfo{volume}{495}}, \bibinfo{pages}{90-97}
  (\bibinfo{year}{2020}).
  
\bibitem{Just2021}
\bibinfo{author}{{Just}, O. et al},
\newblock \bibinfo{title}{{Neutrino absorption and other physics dependencies in neutrino-cooled black hole accretion discs}}.
\newblock \emph{\bibinfo{journal}{Mon. Not. R. Astron. Soc.}} \textbf{\bibinfo{volume}{509}}, \bibinfo{pages}{1377-1412}
  (\bibinfo{year}{2021}).

\bibitem{Ito2021}
\bibinfo{author}{{Ito}, H. et al},
\newblock \bibinfo{title}{{A Global Numerical Model of the Prompt Emission in Short Gamma-ray Bursts}}.
\newblock \emph{\bibinfo{journal}{Astophys. J.}} \textbf{\bibinfo{volume}{918}}, \bibinfo{pages}{59}
  (\bibinfo{year}{2021}).

\bibitem{Just2022}
\bibinfo{author}{{Just}, O. et al},
\newblock \bibinfo{title}{{Dynamical ejecta of neutron star mergers with nucleonic weak processes - II: kilonova emission}}.
\newblock \emph{\bibinfo{journal}{Mon. Not. R. Astron. Soc.}} \textbf{\bibinfo{volume}{510}}, \bibinfo{pages}{2820-2840}
  (\bibinfo{year}{2022}). 

\bibitem{Fujibayashi2018}
\bibinfo{author}{{Fujibayashi}, S. et al},
\newblock \bibinfo{title}{{Mass Ejection from the Remnant of a Binary Neutron Star Merger: Viscous-radiation Hydrodynamics Study}}.
\newblock \emph{\bibinfo{journal}{Mon. Not. R. Astron. Soc.}} \textbf{\bibinfo{volume}{860}}, \bibinfo{pages}{64}
  (\bibinfo{year}{2018}). 
  
\bibitem{Ardevol-Pulpillo2019}
\bibinfo{author}{{Ardevol-Pulpillo}, R. et al},
\newblock \bibinfo{title}{{Improved leakage-equilibration-absorption scheme (ILEAS) for neutrino physics in compact object mergers}}.
\newblock \emph{\bibinfo{journal}{Mon. Not. R. Astron. Soc.}} \textbf{\bibinfo{volume}{485}}, \bibinfo{pages}{4754-5789}
  (\bibinfo{year}{2019}). 

\bibitem{Korobkin2012}
\bibinfo{author}{{Korobkin}, O. et al},
\newblock \bibinfo{title}{{Improved leakage-equilibration-absorption scheme (ILEAS) for neutrino physics in compact object mergers}}.
\newblock \emph{\bibinfo{journal}{Mon. Not. R. Astron. Soc.}} \textbf{\bibinfo{volume}{426}}, \bibinfo{pages}{1940-1949}
  (\bibinfo{year}{2012}). 

\bibitem{Blandford1977}
\bibinfo{author}{{Blandford}, R.~D.} \& \bibinfo{author}{{Znajek}, R.~L.}
\newblock \bibinfo{title}{{Electromagnetic extraction of energy from Kerr black holes}}.
\newblock \emph{\bibinfo{journal}{Mon. Not. R. Astron. Soc.}} \textbf{\bibinfo{volume}{179}},
\bibinfo{pages}{433-456} (\bibinfo{year}{1997}).


\bibitem{Thompson2004}
\bibinfo{author}{{Thompson}, T.~A. et al.}
\newblock \bibinfo{title}{{Magnetar Spin-Down, Hyperenergetic Supernovae, and Gamma-Ray Bursts}}.
\newblock \emph{\bibinfo{journal}{Astrophys. J.}} \textbf{\bibinfo{volume}{611}},
\bibinfo{pages}{380-393} (\bibinfo{year}{2004}).


\bibitem{Metzger2018}
\bibinfo{author}{{Metzger}, B.~D. et al},
\newblock \bibinfo{title}{{A Magnetar Origin for the Kilonova Ejecta in GW170817}}.
\newblock \emph{\bibinfo{journal}{Astophys. J.}} \textbf{\bibinfo{volume}{856}}, \bibinfo{pages}{101}
  (\bibinfo{year}{2018}). 

\end{thebibliography}

\bmhead{Acknowledgements}
We thank Anja C. Andersen, Jonatan Selsing, Radek Wojtak, Katrine Frantzen, Charles Steinhardt, and Christian Vogl for useful discussions. We thank the
ESO Director General for allocating Director’s Discretionary Time to this programme, and the ESO operation staff for support. 
D.W. is supported in part by Independent Research Fund Denmark grant DFF-7014-00017. The Cosmic Dawn Center is funded by the Danish National Research Foundation under grant number 140. A.B. and O.J. acknowledge support from the European Research Council (ERC) under the European Union’s Horizon 2020 research and innovation programme under grant agreement No.\ 759253. A.B. acknowledges support from Deutsche Forschungsgemeinschaft (DFG, German Research Foundation) - Project-ID 279384907 - SFB 1245 and DFG - Project-ID 138713538 - SFB 881 (``The Milky Way System'', subproject A10) and support from the State of Hesse within the Cluster Project ELEMENTS. O.J. acknowledges computational support by the HOKUSAI computer center at RIKEN and by the VIRGO cluster at GSI. R.K. acknowledges support from the Academy of Finland (340613). S.A.S. acknowledges funding from the UKRI STFC Grant ST/T000198/1. D.P. acknowledges support from Israel Science Foundation (ISF) grant 541/17.

\begin{figure}
    \centering
    \includegraphics[width=0.48\linewidth]{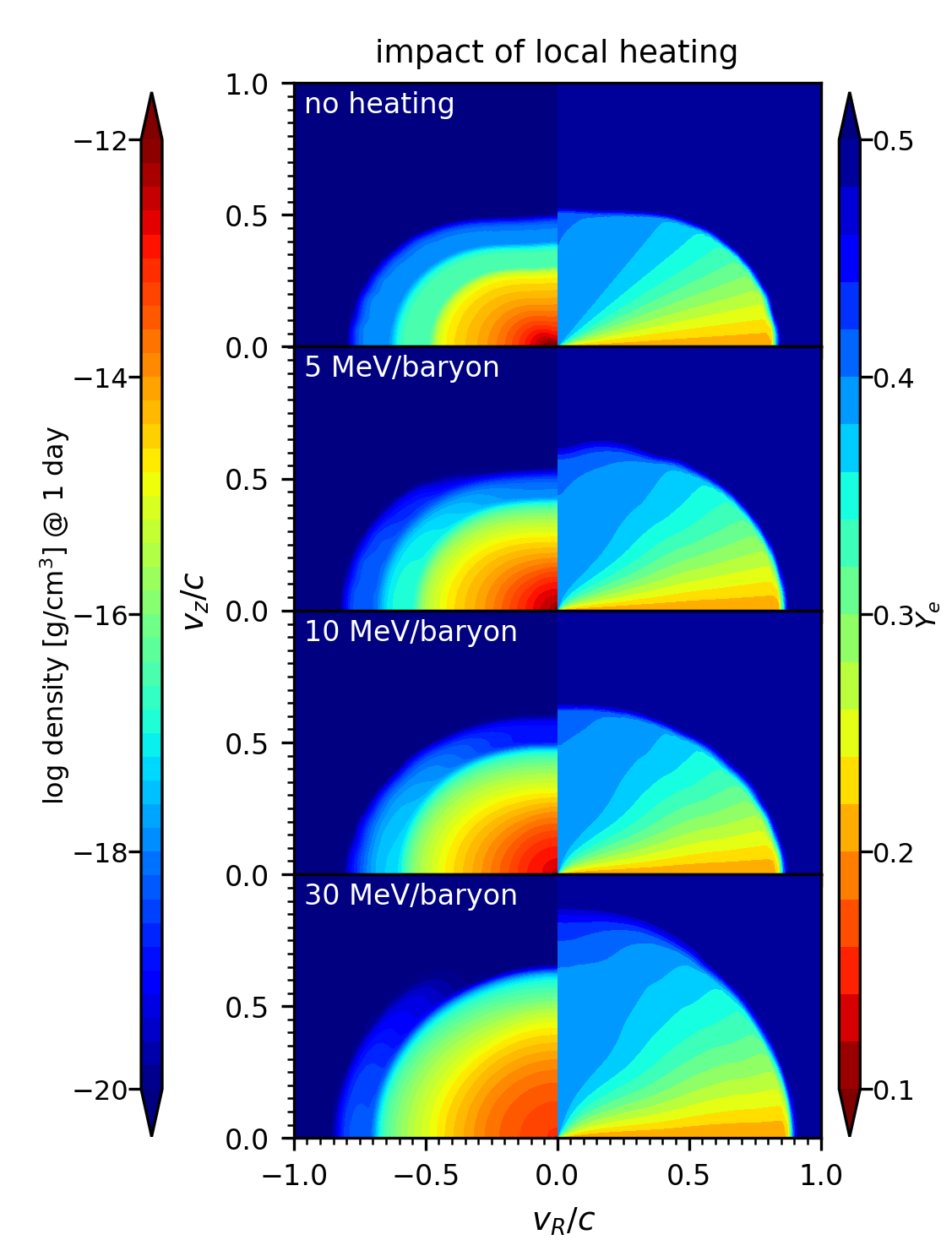}
    \includegraphics[width=0.48\linewidth]{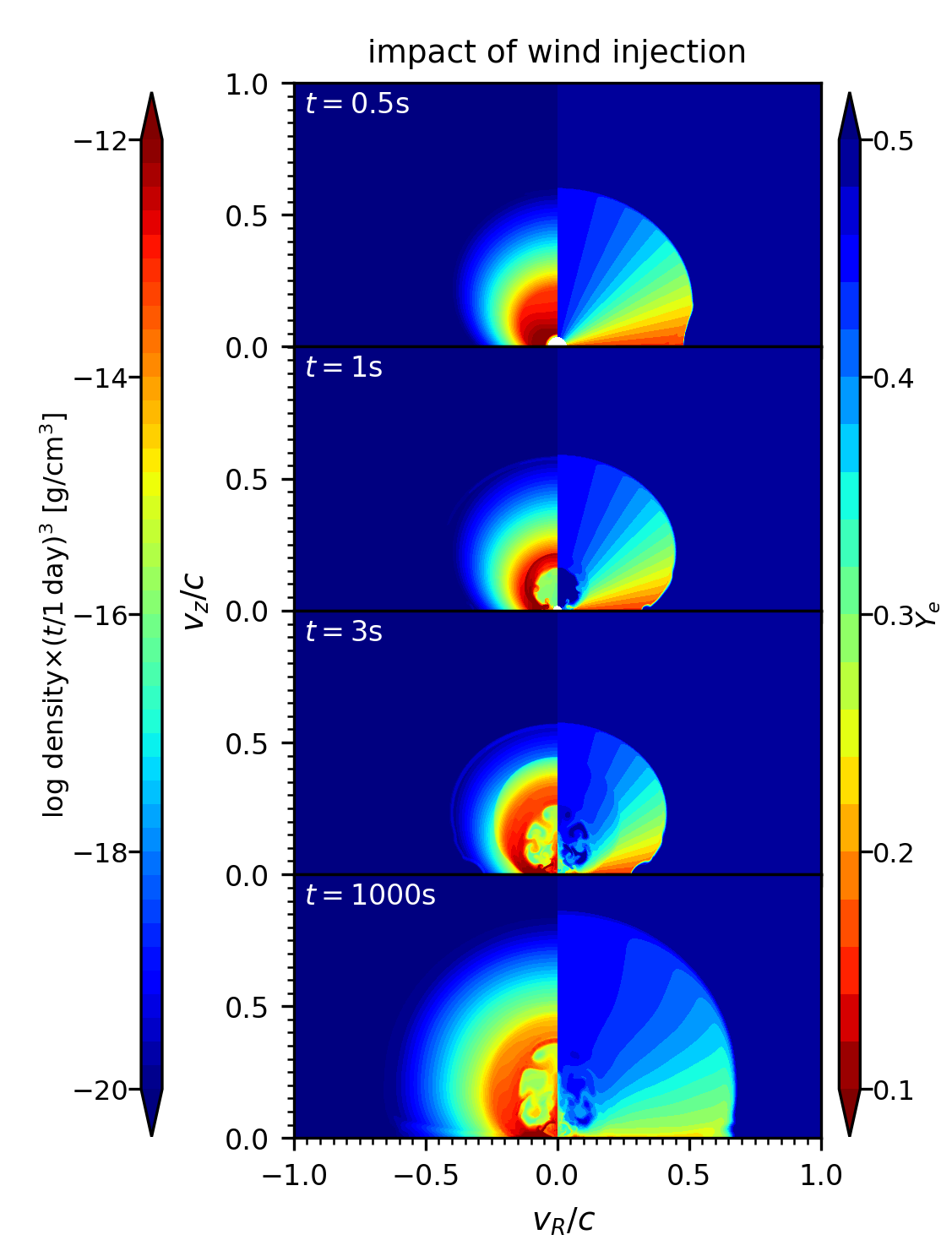}
    \caption{\textbf{Numerical models of energy injection into an expanding cloud of merger ejecta.} The left panel provides color maps of the density (left) and a tracer of the original electron fraction ($Y_e$, right) in velocity space as resulting after 1\,day for four ejecta models in which different amounts of heating energy ($0, 5, 10,$~and~30\,MeV per baryon) were injected during roughly the first second of expansion. While the density distribution can be made spherical with large injection energies, the $Y_e$ stratification remains nearly unchanged.
The right panel shows a model where a relativistic wind with $60^\circ$ half-opening angle around the polar axis is injected. The plots display the same quantities as in the left panels for four different time steps. The wind inflates the innermost part of the ejecta, creating a hot low-density bubble, and launches a shock wave, which dissipates energy predominantly in the polar ejecta, allowing them to spread sideways and, by that, reduce the pole-to-equator variation of $Y_e$. }
    \label{fig:injection}
\end{figure}

\bmhead{Author Contributions}
A.S. and D.W. were the primary drivers of the project and wrote the main text
and developed the figures.  A.S. did all of the analysis and calculations,
wrote most of the Methods sections, and produced Figs.~1--4 and ED1--3.  OJ performed the hydrodynamical simulations and
produced \ref{fig:injection}.  O.J. and A.B. wrote the parts of the main text
and methods sections related to the simulations.  All authors were involved
in interpreting the results and discussed the results and commented on
and/or edited the text.

\bmhead{Competing interests} The authors declare no competing interests.

\newpage

\end{document}